\documentclass[a4paper,11pt]{article}
\usepackage[noblocks]{authblk}
\usepackage{times}
\usepackage{a4wide}

\title{Energy-based Analysis of Biochemical Cycles using Bond Graphs}
 \author{Peter J. Gawthrop}
 \affil{
   Systems Biology Laboratory,
   Melbourne School of Engineering,
   University of Melbourne,
   Victoria 3010, Australia.}

\author{Edmund J. Crampin}
\affil{
   Systems Biology Laboratory,
   Melbourne School of Engineering,
   University of Melbourne,
   Victoria 3010, Australia.
   \authorcr 
   Department of Mathematics and Statistics,
   University of Melbourne,
   Victoria 3010, Australia.
   \authorcr 
   School of Medicine,
   University of Melbourne,
   Victoria 3010, Australia}

\usepackage{amsmath,amssymb,amsbsy,amscd,amstext,mathtools}
\renewcommand{\v}{v}

\newcommand{\ddt}[1]{\dot{#1}}
\newcommand{\dX}{\ddt{X}}
\newcommand{\dx}{\ddt{x}}

\newcommand{\reacul}[2]{
  {\; \xrightleftharpoons[#2]{#1} \;}
}

\newcommand{\reacu}[1]{
  \reacul{#1}{}
}

\newcommand{\reac}{
  \reacu{}
}

\newcommand{\BG}[1]{\text{\sffamily\textbf{#1}}}

\newcommand{\C}{\BG{C }}
\newcommand{\CS}{\BG{CS }}

\newcommand{\one}{\BG{1 }}
\newcommand{\zero}{\BG{0 }}
\newcommand{\One}{\BG{1}}
\newcommand{\Zero}{\BG{0}}

\newcommand{\TF}{\BG{TF }}
\newcommand{\TD}{\BG{TD }}

\renewcommand{\Re}{\BG{Re }}
\newcommand{\mRe}{\BG{mRe }}

\newcommand{\BGL}[2]{$\BG{#1}$:$\mathbf{#2}$} 

\newcommand{\BC}[1]{\BGL{C}{#1}}
\newcommand{\BCS}[1]{\BGL{CS}{#1}}

\newcommand{\BSf}[1]{\BGL{Sf}{#1}}

\newcommand{\BTF}[1]{\BGL{TF}{#1}}

\newcommand{\BRe}[1]{\BGL{Re}{#1}}

\usepackage[numbers,compress]{natbib}
\usepackage{graphicx,subfigure,fancybox,color}

\newcommand{\SubFig}[3]{
 \subfigure[#2]{
   \includegraphics[width=#3\linewidth]{#1.eps}
   \label{subfig:#1}
 }
}

\usepackage[all,dvips]{xy}
\usepackage{array}
\usepackage{color}

\newcommand{\blab}[1]{
{\color{blue}\textstyle #1}
}

\newcommand{\ext}[1]{
  {\color{green}#1}
}

\newcommand{\bd}[1]{
\ar@_{->}[#1]
}

\newcommand{\ebd}[1]{
\ar@[green]@_{->}[#1]
}

\newcommand{\bdn}[3]{
  \bd{#1}^{\blab{#2}}_{\blab{#3}}
}

\newcommand{\bdl}[1]{
\ar@{^|-_{>}}[#1]
\ar@{_|-_{>}}[#1]
}

\newcommand{\bdr}[1]{
\ar@_{->}[#1]
\ar@{-^|}[#1]
\ar@{-_|}[#1]
}

\newcommand{\bdln}[3]{
\ar@{^|-_{>}}[#1]^{\blab{#2}}_{\blab{#3}}
\ar@{_|-_{>}}[#1]
}

\newcommand{\bdrn}[3]{
\ar@_{->}[#1]^{\blab{#2}}_{\blab{#3}}
\ar@{-^|}[#1]
\ar@{-_|}[#1]
}

\newcommand{\burn}[3]{
\ar@^{->}[#1]^{\blab{#2}}_{\blab{#3}}
\ar@{-^|}[#1]
\ar@{-_|}[#1]
}

\newcommand{\bu}[1]{
\ar@^{->}[#1]
}

\newcommand{\ebu}[1]{
\ar@[green]@^{->}[#1]
}

\newcommand{\bun}[3]{
  \bu{#1}^{\color{blue}\textstyle #2}_{\color{blue}\textstyle #3}
}

\newcommand{\ebun}[3]{
  \ebu{#1}^{\color{blue}\textstyle #2}_{\color{blue}\textstyle #3}
}

\newcommand{\bul}[1]{
\ar@{^|-^{>}}[#1]
\ar@{_|-^{>}}[#1]
}

\newcommand{\bur}[1]{
\ar@^{->}[#1]
\ar@{-^|}[#1]
\ar@{-_|}[#1]
}

\newcommand{\buln}[3]{
  \bul{#1}^{\color{blue}\textstyle #2}_{\color{blue}\textstyle #3}
}

\newcommand{\ire}[2]{
\ar@^{>}@<0.3ex>[#1]^{#2}
}

\usepackage[pagebackref]{hyperref}
\hypersetup{colorlinks=true,citecolor=blue,linkcolor=blue,urlcolor=blue}
\usepackage{doi}

\begin{document}
\maketitle
 \begin{abstract}
   Thermodynamic aspects of chemical reactions have a long history in the
   Physical Chemistry literature. In particular, biochemical cycles 
   require a source of energy to function.
   However, although fundamental, the role of chemical potential and
   Gibb's free energy in the analysis of biochemical systems is often
   overlooked leading to models which are physically impossible.

   The bond graph approach was developed for modelling engineering
   systems where energy generation, storage and transmission are
   fundamental. The method focuses on how power flows between
   components and how energy is stored, transmitted or dissipated
   within components.
   Based on early ideas of network thermodynamics, we have applied
   this approach to biochemical systems to generate models which
   automatically obey the laws of thermodynamics. We illustrate the
   method with examples of biochemical cycles.

   We have found that thermodynamically compliant models of simple
   biochemical cycles can easily be developed using this approach. In
   particular, both stoichiometric information and simulation models
   can be developed directly from the bond graph.
   Furthermore, model reduction and approximation while retaining structural and thermodynamic properties is facilitated. 
   Because the bond graph approach is also modular and scaleable, we
   believe that it provides a secure foundation for building
   thermodynamically compliant models of large biochemical networks.
 \end{abstract}
\newpage
\section{Introduction}
\label{sec:intro}
\begin{quotation}
   \noindent\emph{Oh ye seekers after perpetual motion, how many vain chimeras
   have you pursued? Go and take your place with the alchemists.}
   Leonardo da Vinci, 1494
\end{quotation}

Thermodynamic aspects of chemical reactions have a long history in the
Physical Chemistry literature. In particular, the role of chemical
potential and Gibb's free energy in the analysis of biochemical
systems is developed in, for example, the textbooks of \citet{Hil89},
\citet{BeaQia10} and \citet{KeeSne09}.
As discussed by, for example, \citet{KatCur65} and \citet[Chapter
8]{Cel91}, there is a distinction between classical thermodynamics
which treats closed systems which are in equilibrium or undergoing
reversible processes and non-equilibrium thermodynamics which treats
systems, such as living organisms, which are open and irreversible.

Biochemical cycles are the building-blocks of biochemical systems; as
discussed by \citet{Hil89}, they require a source of energy to
function. For this reason, the modelling of biochemical cycles
requires close attention to thermodynamical principles 
to avoid models which are physically impossible. Such physically
impossible models are analogous to the perpetual motion machines
beloved of inventors.
In the context of biochemistry, irreversible reactions are \emph{not},
in general, thermodynamically feasible and can be erroneously used to
move chemical species against a chemical gradient thus generating
energy from nothing~\citep{Gun14}. 
%
The theme of this paper is that models of biochemical networks must
obey the laws of thermodynamics; therefore it is highly desirable to
specify a modelling framework in which compliance with thermodynamic
principles is automatically satisfied. Bond graphs provide one such
framework.
%

Bond graphs were introduced by Henry Paynter (see \citet{Pay92} for a
history) as a method of representing and understanding complex
multi-domain engineering systems such as hydroelectric power
generation. A comprehensive account of bond graphs is given in the
textbooks of \citet{GawSmi96}, \citet{Bor11} and \citet{KarMarRos12} 
 and a tutorial introduction for control
engineers is given by \citet{GawBev07}. 

As discussed in the textbooks of, for example, \citet{Pal06,Pal11},
\citet{Alo07} and \citet{KliLieWie11}, the numerous biochemical
reactions occurring in cellular systems can be comprehended by
arranging them into \emph{networks} and analysing them by graph theory
and using the associated connection matrices.
These two aspects of biochemical reactions -- thermodynamics and
networks -- were brought together some time ago by
\citet{OstPerKat71}. A comprehensive account of the resulting
\emph{network thermodynamics} is given by \citet{OstPerKat73}. As
discussed by \citet{OstPer74} such thermodynamic networks can be
analysed using an equivalent electrical circuit representation; but,
more generally, the bond graph approach provides a natural
representation for network thermodynamics
\citep{OstAus71a,OstAus71b,OstPerKat73}.
This approach was not widely adopted by the biological and biochemical
modelling community, and may be considered to have been ahead of its
time. 
Mathematical modelling and computational analysis of biochemical
systems has developed a great deal since then, and now underpins the
new disciplines of systems biology \citep{Kohl:2010iw}, and
``physiome'' modelling of physiological systems
\citep{AldBurLauSor06,Smith:2007co,Hunter:2008cb,Hunter:2012dd}, where
we are faced with the need for physically feasible models across
spatial and temporal scales of biological organisation.

In particular there has been a resurgence of interest in this approach
to modelling as it imposes extra constraints on models, reducing the
space of possible model structures or solutions for
consideration. This has been applied from individual enzymes
\citep{SmiCra04,Tran:2009ui} and cellular pathways
\citep{Beard:2005uk} up to large scale models
\citep{Beard:2004iu,BeaLiaQia02,Feist:2007dq}, as a way of eliminating
thermodynamically infeasible models of biochemical processes and
energetically impossible solutions from large scale biochemical
network models alike (see \citet{Soh:2010je} for a review).
Additionally, there is new impetus into model sharing and reuse in the
biochemical and physiome modelling communities which has garnered
interest in modular representations of biochemical networks, and has
promoted development of software, languages and standards and
databases for models of biochemical processes. Model representation
languages such as CellML  and SBML
 promote model sharing through databases such as
the Physiome Model Repository  and BioModels Database. Descriptions of models in a hierarchical and
modular format allows components of models to be stored in such
databases and assembled into new models.
Rather than to revisit the detailed theoretical development, therefore, our aim is therefore to refocus attention on the bond graph representation of biochemical networks for practical purposes such as these. 
First we briefly review the utility of the bond graph approach with these aims in mind. 


Bond graph approaches have also developed considerably in recent
years, in particular through the development of computational tools
for their analysis, graphical construction and manipulation, and
modularity and reuse
\citep{BalBevGawDis05,CelNeb05,Bor06,CelGre08,CelGre09,CalCelYebDor13}, which are key
preoccupations for systems biology and physiome modelling. Our focus
is on how kinetics and thermodynamic properties of biochemical
reactions can be represented in this framework, and how the bond graph
formalism allows key properties to be calculated from this
representation.
In addition, bond graph approaches have been extended in recent years
to model electrochemical storage devices \citep{Kar90} and heat
transfer in the context of chemical reactions
\citep{ThoAtl77}. \citet{Cel91} extends network
thermodynamics beyond the isothermal, isobaric context of
\citet{OstPerKat73} by accounting for both work and heat and a series
of papers \citep{CelGre09,GreCel12} shows how multi-bonds can be used
to model the thermodynamics of chemical systems with heat and work
transfer and convection and to simulate large systems.
\citet{ThoAtl85} discuss ``osmosis as chemical
reaction through a membrane''. \citet{LefLefCou99} model cardiac
muscle using the bond graph approach.



Bond graphs explicitly model the flow of energy through networks making
use of the concept of power covariables: pairs of variable whose
product is power. For example, in the case of electrical
networks, the covariables are chosen as voltage and current.  As
discussed by a number of authors \citep{Hil89,Cel91,Fuc96,JobHer06},
chemical potential is the driving force of chemical reactions. Hence,
as discussed by \citet{Cel91}, the appropriate choice of power
covariables for isothermal, isobaric chemical reaction networks is
chemical potential and molar flow rates.
As pointed out by \citet{BeaLiaQia02}, using both mass and energy
balance ensures that models of biochemical networks are
thermodynamically feasible.  Modelling using bond graphs automatically
ensures not only mass-balance but also energy-balance; thus models of
biochemical networks developed using bond graphs are thermodynamically
feasible. 

As discussed by \citet{Hil89}, biochemical cycles are the
building-blocks of biochemical systems. 
Bond graph models are able to represent thermodynamic cycles
and therefore appropriately represent free energy transduction in 
biochemical processes in living systems.


Living systems are complex, and therefore a hierarchical and modular approach
to modelling biochemical systems is desirable.  
Bond graphs have a natural hierarchical
representation \citep{Cel92} and have been used to model complex network
thermodynamics \citep{Cel91,CelGre09,GreCel12}. 
%
Complex systems can be simplified by approximation: the bond graph
method has a formal approach to
approximation~\citep{GawSmi96,GawBev07,KarMarRos12} and the potential algebraic
issues arising from such approximation~\citep{GawSmi92a}. In
particular, complex systems can be simplified if they exhibit a fast
and slow timescale; a common feature of many biochemical (for example,
Michaelis-Menten enzyme kinetics) and cell physiological systems (for
example, slow-fast analysis of the membrane potential of electrically
excitable cells).  A bond graph approach to two-timescale
approximation has been presented by \citet{SueDau91a}.



As well as providing a thermodynamically-consistent model of a
dynamical system suitable for simulation, representation of a
biochemical system using the bond graph approach enables a wide range
of properties and characteristics of the system to be represented.
A number of key physical properties can be derived directly from the
bond graph representation. For example, chemical reactions involve
interactions between species which preserve matter; the number of
moles of each species in a reaction must be accounted for. 


As discussed by \citet[\S5.2]{OstPerKat73}, the kinetics of
biochemical networks become particularly simple near thermodynamic
equilibrium. However, as discussed by \citet{QiaBea05}, it is
important to consider the behaviour of biochemical networks in living
systems far from equilibrium. In particular, the analysis of
non-equilibrium steady-states (where flows are constant but non-zero
and states are constant) is important
\citep{QiaBeaLia03,BeaQia10}.


Using elementary reactions as examples, \S\ref{sec:bg} shows
how biochemical networks may be modelled using bond graphs. 
The bond graph is more than a sketch of a biochemical network; it
can be directly interpreted by a computer and, moreover, has a number
of features that enable key physical properties to be derived from the
bond graph itself. For example, \S\ref{sec:causality} shows how
the bond graph 
can be used to examine the
\emph{stoichiometric} properties of biochemical networks.
%
%
%
\S\ref{sec:approximation} discusses the role of bond graphs in
the structural approximation of biochemical networks.
%
\S\ref{sec:cycles} discusses two biochemical cycles, an enzyme
catalysed reaction and a biochemical switch, to
illustrate the main points of the paper.
%
\S\ref{sec:hierarchical} discusses software aspects of the Bond
Graph approach and how it could be integrated into preexisting
hierarchical modelling frameworks.
\S\ref{sec:conclusion} concludes the paper.
%


\section{Bond Graph  Modelling of Chemical Reactions}
\label{sec:bg}
\begin{figure}[htbp]
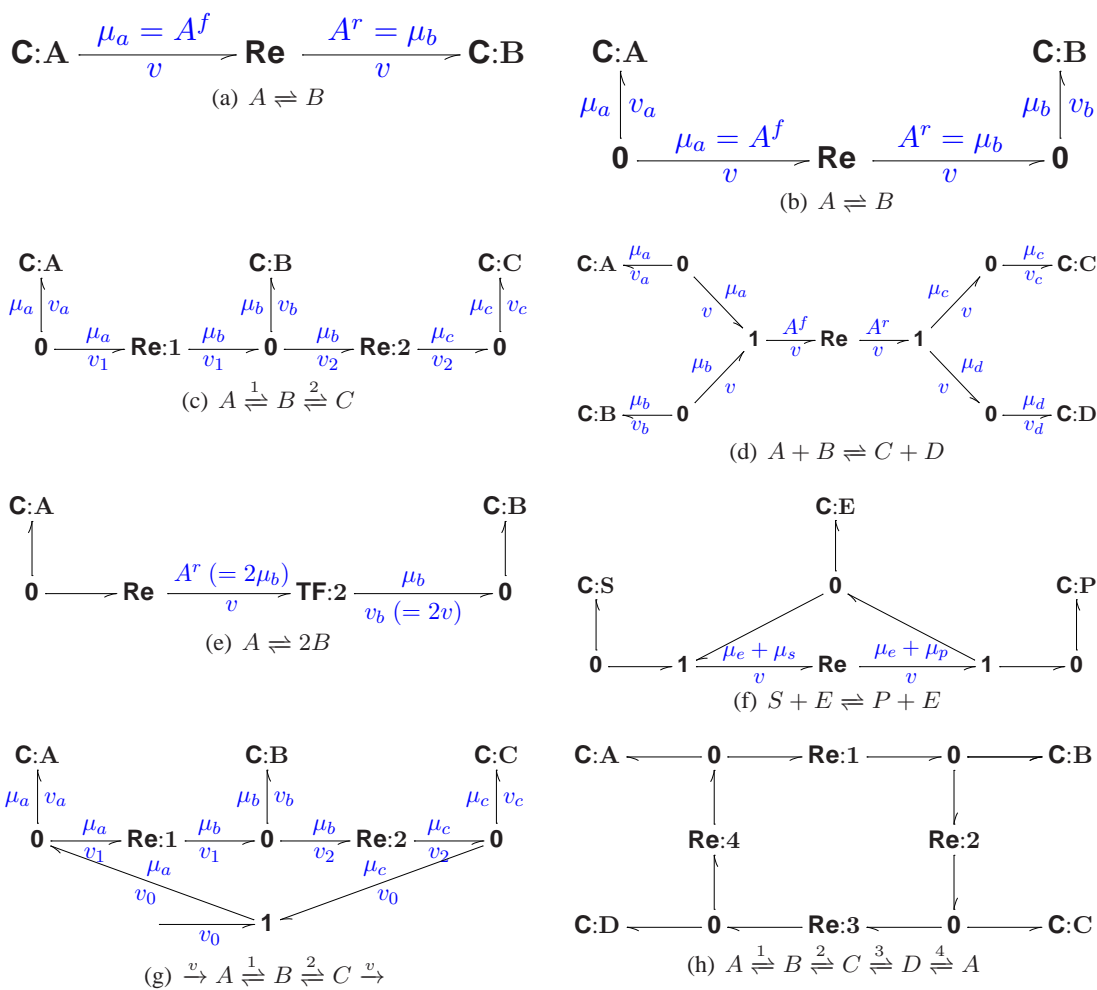

 \centering
 
 \subfigure[$A \reac B$]{
   \resizebox{0.45\linewidth}{!}{\input{AB_bg}}
   \label{subfig:AB_bg}
 }

 \subfigure[$A \reac B$]{
   \resizebox{0.45\linewidth}{!}{\input{AB_1_bg}}
   \label{subfig:AB_1_bg}
 }

 \subfigure[$A \reacu{1} B \reacu{2} C$]{
   \resizebox{0.45\linewidth}{!}{\input{ABC_bg}}
   \label{subfig:ABC_bg}
 }

 \subfigure[$A+B \reac C+D $]{
   \resizebox{0.45\linewidth}{!}{\input{ABCD_bg}}
   \label{subfig:ABCD_bg}
 }

 \subfigure[$A \reac 2B$]{
   \resizebox{0.45\linewidth}{!}{\input{A2B_ec_bg}}
   \label{subfig:A2B_ec_bg}
 }

 \subfigure[$S+E \reac P+E$]{
   \resizebox{0.45\linewidth}{!}{\input{ECR_bg}}
   \label{subfig:ECR_bg}
 }

 \subfigure[$ \xrightarrow{v} A \reacu{1} B\reacu{2} C\xrightarrow{v}$]{
   \resizebox{0.45\linewidth}{!}{\input{ABCv_bg}}
   \label{subfig:ABCv_bg}
 }

 \subfigure[$A \reacu{1} B \reacu{2} C \reacu{3} D \reacu{4} A$]{
   \resizebox{0.45\linewidth}{!}{\input{Cycle_bg}}
   \label{subfig:Cycle_bg}
 }

 \caption[Simple reactions and their bond graphs]
 {Simple reactions and their bond graphs. (a) The simple binary
   reaction is represented by a bond graph using a \C component for
   each substance and an \Re component to explicitly represent the
   reaction. (b) An alternative representation using \Zero~(common
   potential) junctions to allow connections. (c) Two reactions in
   series extending (b). (d) A single reaction between four substances
   requires a single \Re component, one \C component for each
   substance and two \One~(common flow) connections. (e) The
   stoichiometric coefficient $2$ can be incorporated using the bond
   graph \TF~component. (f) A simple enzyme-catalysed reaction. The
   enzyme $E$ appears on each side of the formula thus creating a
   cycle in the bond graph (see \S\ref{sec:cycles}\ref{sec:MM}). (g)
   The same as (c) but with an externally-imposed flow that adds
   molecules of $A$ whilst subtracting the same number of molecules of
   $C$ thus allowing a non-equilibrium steady-state. (h) A simple
   biochemical cycle.}
 \label{fig:simple}
\end{figure}

Bond graphs are an energy-based modelling approach. This section
introduces the bond graph methodology in the context of biochemical
reactions using the reactions listed in figure
\ref{fig:simple}. The section is organised to
emphasise the key aspects of bond graph modelling which make it a
powerful approach to the modelling of biochemical systems.


\subsection{Energy flow, storage and dissipation in a simple reversible reaction}
\label{sec:energy}
figure
\ref{subfig:AB_bg} shows the simple interconversion of two molecular species, A and B. As mentioned above, a thermodynamically consistent representation of biochemical processes demands consideration of reversible reactions, and so we consider this simple interconversion as the simplest possible reaction. This interconversion is represented by \emph{bonds} of the form
$\xymatrix{\bdn{r}{\mu}{v}&}$
each of which is associated with two variables\footnote{The textual
 annotation in blue is for explanatory purposes, it is not part of
 the bond graph itself.}:
the \emph{chemical potential} $\mu$ $(\text{J mol}^{-1})$ and a
\emph{molar flow rate} $v$ $(\text{mol s}^{-1})$\footnote{The standard
 bond graph terminology is that the chemical potential is termed an
 \emph{effort} and is analogous to voltage in electrical systems and
 force in mechanical systems.  Similarly, the molar flow rate is
 termed a \emph{flow} and analogous to current in electrical systems
 and velocity in mechanical systems.}. The product of these two
variables is energy flow or \emph{power} $P = \mu \times v$
$(\text{W})$.
The bonds represent the transmission of power in the system, and do not not create, store or dissipate power. 
The half-arrow on the bond indicates the direction in which power flow
will be regarded as positive and thus defines a sign
convention.

In figure \ref{subfig:AB_bg} the pools of chemical species A and B are
represented by \C components. These components reflect the amount of
each species present (and hence determine the chemical potential of
each species)\footnote{The \C component stands for `Capacitor'. The
  chemical potential is analogous to the voltage associated with a
  capacitor in an electrical circuit, which charges or discharges if
  there is a net influx or efflux into the component.}.
%
\BC{A} contains $x_a$ moles of species $A$ and the rate of decrease
is equal to the molar flow $v$; \BC{B} contains $x_b$ moles of species $B$ and the rate of increase
is $v$. Thus:
\begin{xalignat}{2}
 \dx_a &= -v &
 \dx_b &= v \label{eq:dx_b}
\end{xalignat}
Each \C component is associated with a chemical potential $\mu$
which, assuming a dilute system within a volume $V$, is given by \citep[\S1.2]{KeeSne09}:
\begin{xalignat}{2}
 \mu_a &= \mu_a^0 + RT\ln \frac{x_a}{V}&
 \mu_b &= \mu_b^0 + RT\ln \frac{x_b}{V}\label{eq:mu_a0}
\end{xalignat}
where $\mu_a^0$ is the standard chemical potential for species A, and similarly for species B. 
It is convenient to rewrite Equations (\ref{eq:mu_a0}) as:
\begin{xalignat}{4}
 \mu_a &=  RT\ln K_a x_a &
 \mu_b &= RT\ln K_b x_b &
\text{where } K_a &= \frac{1}{V}e^{\frac{\mu_a^0}{RT}} &
\text{and } K_b &= \frac{1}{V}e^{\frac{\mu_b^0}{RT}}\label{eq:mu_a}
\end{xalignat}
Each \C component \emph{stores} but does not create or dissipate energy. The
corresponding energy flow is described through the bond to which it is connected.

The reversible reaction between chemical species $A$ and $B$ is
represented by a single \Re (Reaction) component which relates the
reaction flow $v$ to the chemical affinities (weighted sum of chemical
potentials) for the forwards and reverse reactions
$A^f=\mu_a$ and $A^r=\mu_b$. 
As discussed by \citet{Rys58} and \citet[\S5.1]{OstPerKat73}, 
the reaction rate, or molar flow, is given by the \emph{Marcelin -- de
 Donder} formula:
\begin{xalignat}{3}
 v &=  v^+ - v^-&
   \text{where }
   v^+ &=  \kappa e^\frac{A^f}{RT}&
   \text{and }
   v^- &=  \kappa e^\frac{A^r}{RT}\label{eq:v_exp}
\end{xalignat}
where $\kappa$ is a constant which determines reaction rate. 
This can be rewritten in two ways. The \emph{de Donder} formula
\citep[Equation(11)]{Bou83}:
\begin{xalignat}{2}
 \frac{v^+}{v^-} &= e^\frac{A}{RT} &
 \text{where }
 A &= A^f - A^r \label{eq:deDonder}
\end{xalignat}
and the \emph{Marcelin} formula \citep[Equation(1)]{Lai85}:
\begin{equation}
 \label{eq:Marcelin}
 v = \kappa 
 \left ( e^\frac{A^f}{RT} -
   e^\frac{A^r}{RT} \right )
\end{equation}
This latter formulation is used in the sequel.
%
The \Re component \emph{dissipates}, but does not create or store, energy. 

In the particular case of figure \ref{subfig:AB_bg}, substituting the
chemical potentials of Equations (\ref{eq:mu_a})
into Equations (\ref{eq:v_exp}) recovers the well
known first-order mass-action expressions:
  \begin{xalignat}{3}
  v^+  &= \kappa K_a x_a&
  v^-  &= \kappa K_b x_b&
  v &=  \kappa  \left( 
    K_a x_a - K_b x_b
  \right)  \label{eq:v_ex_a_0}
  \end{xalignat}
  We note that this notation clearly demarcates parameters relating to
  thermodynamic quantities ($K_a, K_b$) from reaction kinetics
  ($\kappa$) and that the equilibrium constant is given by $K_b/K_a$.

  The Equations \eqref{eq:v_ex_a_0} can also be written in the
  conventional rate constant form as
\begin{xalignat}{3}
 v^+  &=  k^+ x_a&
 v^-  &= k^- x_b&
 v &=  v^+ - v^- = 
   k^+ x_a - k^- x_b \label{eq:v_ex_a}
\end{xalignat}
where the forwards and backwards first order rate constants are 
\begin{equation}\label{eq:rate_constants}
k^+ = \kappa K_a \text{ and } k^- = \kappa K_b
\end{equation}
The thermodynamic quantities and reaction kinetics are no longer
distinguished in the rate constant formulation of Equations
\eqref{eq:v_ex_a}.

\subsection{Modularity: coupling reactions into networks}
\label{sec:modularity}
A key feature of bond graph representations is to construct and analyse models of large scale systems from simpler building blocks.  
The bond graph of figure \ref{subfig:AB_bg} cannot be used as a
building block of a larger system as there are no connections
available with which to couple to other reactions. However, the bond graph approach is, in general, modular and
provides two connection components for this purpose: the
\zero junction and the \one junction. Each of these components
transmits, but does not store, create or dissipate energy. 
In figure \ref{subfig:AB_1_bg} the representation of the simple
reversible reaction in figure \ref{subfig:AB_bg} is expanded to
include two \zero junction connectors. This representation is identical to that in
figure \ref{subfig:AB_bg} except that it makes explicit the junctions
through which other reactions involving species A and B can be coupled
to this reaction.
The bond graph of figure \ref{subfig:ABC_bg} makes use of the
right-hand \zero junction of figure \ref{subfig:AB_1_bg} to build two
connected reactions; where species B is also reversibly interconverts
with species C.

The connector in this case is a \zero junction. 
The \zero junction can have two or more impinging bonds. In the case
of the central \zero junction  of  figure \ref{subfig:ABC_bg}, there
are three impinging bonds: one
($\xymatrix{\bdn{r}{\mu_b}{v_1}&}$)
pointing in and two
($\xymatrix{\bdn{r}{\mu_b}{v_b}&}$ and $\xymatrix{\bdn{r}{\mu_b}{v_2}&}$)
pointing out. As indicated in figure \ref{subfig:ABC_bg}, the \zero
junction has two properties:
\begin{enumerate}
\item the chemical potentials or affinities (\emph{efforts}) on all impinging
 bonds are constrained to be the same, (the \zero junction is therefore a \emph{common potential} connector), and
\item the molar \emph{flows} on the impinging bonds sum to zero, under the sign convention that a plus
 sign is appended to the flows corresponding to inward bonds and a
 minus sign for outward bonds:
 \begin{xalignat}{2}
   v_1 - v_b - v_2 &= 0&
   \text{or } 
   v_b &= v_1 - v_2
 \end{xalignat}
\end{enumerate}
These two properties imply a third: the power flowing out of a \zero
junction is equal to the power flowing in (the \zero junction is power-conserving):
\begin{equation}
 P_a + P_2 = \mu_bv_b + \mu_bv_2
 = \mu_b (v_b + v_2)
 = \mu_b v_1
 = P_1
\end{equation}
In a similar fashion, the left-hand \zero junction implies that $v_a = -v_1$
and the right-hand \zero junction implies that $v_c = v_2$. 
Figure \ref{subfig:ABC_bg} can easily be extended to give a reaction
chain of arbitrary length.

In contrast, in order to represent the reaction of figure
\ref{subfig:ABCD_bg} we introduce the \one junction, which has the same
power-conserving property as the \zero junction but which represents a
\emph{common flow} connector\footnote{The common flow \one junction is
 the \emph{dual} component of the common effort \zero junction.}.
In particular, with reference to the left-hand
\one junction in figure \ref{subfig:ABCD_bg}:
\begin{enumerate}
\item the \emph{molar flows} on all impinging bonds are constrained to be
 the same and
\item the \emph{affinities} on the impinging bonds sum to zero when a
 plus sign is appended to the efforts corresponding to inward bonds
 and a minus sign for outward bonds:
 \begin{xalignat}{2}
   \mu_a + \mu_b - A^f &= 0&
   \text{or } 
   A^f &= \mu_a + \mu_b \label{eq:c_A_f}
 \end{xalignat}
\end{enumerate} 
Similarly, the right-hand \one junction implies that:
\begin{xalignat}{2}
 A^r - \mu_c - \mu_d  &= 0&
 \text{or } 
 A^r &= \mu_c + \mu_d \label{eq:c_A_r}
\end{xalignat}
substituting the chemical potentials of Equations (\ref{eq:c_A_f}) and
(\ref{eq:c_A_r}) into Equations (\ref{eq:v_exp})
gives the well known second-order mass-action expression:
\begin{align}
 v &= \kappa  \left( 
   K_a x_a K_b x_b - K_c x_c K_d x_e 
 \right)
 = k^+ x_a x_b - k^- x_c x_d\\
\text{where }
 k^+ &= \kappa K_a K_b 
\text{ and }
 k^- = \kappa K_c K_d \label{eq:v_ex_c}
\end{align}
Once again, we note that this notation clearly demarcates parameters
relating to thermodynamic quantities ($K_a, K_b, K_c, K_d$) from reaction
kinetics ($\kappa$).
\subsection{Incorporating stoichiometry into reactions}
\label{sec:stoich}
The reaction of figure \ref{subfig:A2B_ec_bg} has one mole of species $A$
reacting to form two moles of species $B$. The corresponding bond
graph uses the \TF\footnote{\citet{OstPerKat73} use the symbol
 \TD in place of \TF.} component to represent this stoichiometry. The \TF
component transmits, but does not store, create or dissipate energy.
Hence, the power out equals the power in. Thus in the context of
figure \ref{subfig:A2B_ec_bg}:
\begin{equation}
 \label{eq:TF}
 A^r v = \mu_b v_b
\end{equation}
(noting that in this case $A^r$ is the `unknown' as $\mu_b$ is determined by the \zero junction). 

A \TF component with ratio $n$ is donated by \BTF{n} and is defined by
the power conserving property and that the output flow is $n$ times the input
flow. As power is conserved, it follows therefore that the input effort is $n$ times the output effort. 
In the context of figure \ref{subfig:A2B_ec_bg}:
\begin{xalignat}{2}
 A^r &= 2 \mu_b&
 v_b &= 2 v
\end{xalignat}
Noting that $A_f = \mu_a$ it follows from Equations (\ref{eq:v_exp}) that:
\begin{align}
 v &= \kappa  \left( 
   K_a x_a  - (K_b x_b)^2
 \right)
 = k^+ x_a  - k^- x_b^2   \label{eq:v_ex_d}
 \text{ where }
 k^+ = \kappa K_a 
 \text{ and }
 k^- = \kappa K_b^2
\end{align}

\subsection{Non-equilibrium steady states: reactions with external flows.}
\label{sec:ext_flow}
As has been discussed by many authors, in cells biochemical reactions are maintained away from thermodynamic equilibrium through continual mass and energy flow through the reaction. 
The reaction of figure \ref{subfig:ABCv_bg} corresponds to the
reaction in Figure \ref{subfig:ABC_bg} except that an external flow
$v_0>0$ has been included. This corresponds to adding molecules of
$A$ and removing molecules of $C$ at the same fixed rate. As
discussed by \citet{QiaBeaLia03}, the reaction has a non-equilibrium
steady-state (NESS) corresponding to $v_1=v_2=v_0$. This is a
steady-state because the flows $v_a=v_b=v_c=0$ and hence
$\dx_a=\dx_b=\dx_c=0$; it is not a thermodynamic equilibrium because
$v_1\ne 0$ and $v_2 \ne 0$.

\subsection{Thermodynamic compliance}
\label{sec:compliance}
The bond graph approach ensures thermodynamic compliance: the
model may not be correct, but it does obey the laws of thermodynamics.
To illustrate this point, consider the Biochemical Cycle of figure
\ref{subfig:Cycle_bg}. As discussed by, for example,
\citet{QiaBeaLia03}, a fundamental property of such cycles is the
thermodynamic constraint that
\begin{equation}\label{eq:constraint}
\frac{k_{+1}k_{+2}k_{+3}k_{+4}}{k_{-1}k_{-2}k_{-3}k_{-4}} = 1
\end{equation}
This property arises from the requirement for detailed balance around the biochemical cycle. 
However, as is now shown, the thermodynamic constraint of Equation
(\ref{eq:constraint}) is automatically satisfied by the bond graph
representation of figure \ref{subfig:Cycle_bg}.

In the same way as Equation (\ref{eq:v_ex_a}), the four reaction flows
can be written as:
\begin{xalignat}{3}
v_1 &=  \kappa_1  \left( K_a x_a - K_b x_b \right), &
&\dots,&
v_4 &=  \kappa_4  \left( K_d x_d - K_a x_a \right) \label{eq:cons_v_1}
\end{xalignat}
Alternatively, the four reaction flows of Equations
(\ref{eq:cons_v_1}) can be
rewritten as:
\begin{xalignat}{3}
v_1 &=   \left( k_{+1} x_a - k_{-1} x_b \right),&
&\dots,&
v_4 &=   \left( k_{+4} x_d - k_{-4} x_a \right) \label{eq:cons_v_1_alt}
\end{xalignat}
where
\begin{xalignat}{6}
k_{+1} &= \kappa_1 K_a,&
&\dots,&
k_{+4} &= \kappa_4 K_d,&
k_{-1} &= \kappa_1 K_b,&
&\dots,&
k_{-4} &= \kappa_4 K_a
\end{xalignat}
Hence:
\begin{equation}\label{eq:const_actual}
  \frac{k_{+1}k_{+2}k_{+3}k_{+4}}{k_{-1}k_{-2}k_{-3}k_{-4}} = 
  \frac{\kappa_1 K_a \kappa_2 K_b \kappa_3 K_c \kappa_4 K_d}
       {\kappa_1 K_b \kappa_2 K_c \kappa_3 K_d \kappa_4 K_a}
\end{equation}
As each factor of the numerator on the right-hand side of Equation
(\ref{eq:const_actual}) appears in the denominator, and \emph{vice
 versa}, then Equation (\ref{eq:constraint}) is satisfied.

\section{Stoichiometric Analysis of Reaction Networks}
\label{sec:causality}
Stoichiometric analysis is fundamental to understanding the properties
of large networks \citep{Pal06, Pal11, JamPal11}.  In particular
computing the left and right null space matrices leads to information
about pools and steady-state pathways
\citep{SchLetPal00,SchHilWooFel02,FamPal03,FamPal03a}.
For example, when analysing reaction networks such as metabolic
networks, one may seek to determine for measured rates of change of
metabolite concentrations, what are the reaction rates in the
network. This question is addressed below.
Initially we will address the inverse problem: for given reaction velocities, what are the rates of change of concentrations of the chemical species?
In bond graph terms, this asks the question: ``given the reaction flows $V$, what are the flows $\dX$ at the \C
components?''. 
This can be addressed directly from the bond graph using the concept of \emph{causality}. 

The bond graph concept of causality~\citep{GawSmi96,GawBev07,KarMarRos12} has proved useful
for generating simulation code, detecting modelling inconsistencies,
solving algebraic loops~\citep{GawSmi92a}, approximation,
inversion~\citep{Gaw00d,NgwScaTho01b,MarJar11} and analysis of system
properties~\citep{SueDau89}.
This section shows how the bond graph concept of causality can be used
to examine the stoichiometry of networks of biochemical reactions. 
As in \S\ref{sec:bg}, this is done by analysis of particular
examples. However, as discussed in Section \ref{sec:hierarchical},
this approach scales up to arbitrarily large systems.

\subsection{The stoichiometric matrix}
\label{sec:integral}
\begin{figure}[htbp]
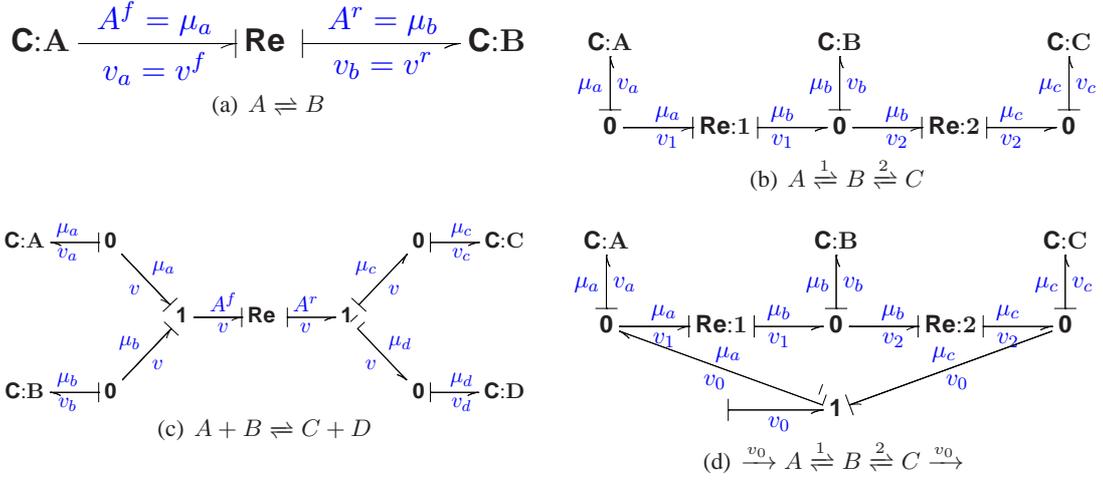

 \centering
 
 \subfigure[$A \reac B$]{
   \resizebox{0.45\linewidth}{!}{\input{AB_icbg}}
   \label{subfig:AB_icbg}
 }

 \subfigure[$A \reacu{1} B \reacu{2} C$]{
   \resizebox{0.45\linewidth}{!}{\input{ABC_icbg}}
   \label{subfig:ABC_icbg}
 }

 \subfigure[$A+B \reac C+D $]{
   \resizebox{0.45\linewidth}{!}{\input{ABCD_icbg}}
   \label{subfig:ABCD_icbg}
 }

 \subfigure[$ \xrightarrow{v_0} A \reacu{1} B\reacu{2} C\xrightarrow{v_0}$]{
   \resizebox{0.45\linewidth}{!}{\input{ABCv_icbg}}
   \label{subfig:ABCv_icbg}
 }

 \caption[Causal strokes and the Stoichiometric Matrix]
 {Causal strokes and the Stoichiometric Matrix. The bond graph notion
   of causality provides an algorithm for determining the
   stoichiometric matrix by explicitly showing how the \Re flows
   propagate to the \C flows.
   (a) The \C components impose a potential onto the \Re component;
   the \Re component imposes a flow into the \C components. 
   (b) As (a) and note that exactly one bond imposes a potential on
   to each \Zero~(common potential) junction. 
   (c) As (a) and note that exactly one bond imposes a flow on to
   each \One~(common flow) junction. 
   (d) As the external flow $v_0$ impinges on to \Zero~junctions, it
   does not affect the causality of the parts in common with (b).
}
 \label{fig:simple-ic}
\end{figure}
Figure \ref{fig:simple-ic}(a) is similar to the bond graph of
figure \ref{subfig:AB_bg} except that two lines have been added
perpendicular to each bond; these lines are called \emph{causal
 strokes}.
It is convenient to distinguish between the flows on each side of the \Re
component by relabelling them as $v^f$ and $v^r$ ($v^f=v_r=v$) and this
is reflected in the annotation.
The implications of the causal stroke are twofold:
\begin{enumerate}
\item The bond imposes \emph{effort} on the component at the stroke
 end of the bond.
\item The bond imposes \emph{flow} on the component at the other
 end of the bond.
\end{enumerate}
Thus, as indicated on the bond graph\footnote{Although in mathematics
 $x=y$, $y=x$ and $x-y=0$ are the same, this is not true in
 imperative programming languages; the left-hand side is computed
 from the right hand side. This latter interpretation is used in the
 rest of this section.}:
the \emph{flows} are given by:
\begin{xalignat}{2}
 \dx_a &= -v_a = -v^f = -v&
 \dx_b &= v_b = v^r = v
\end{xalignat}
and the \emph{efforts} by
\begin{xalignat}{2}
 A_f& = \mu_a&
 A_r &= \mu_b
\end{xalignat}
In general, the reaction flows can be composed into the vector $V$, and
the state derivatives into the vector $X$, and these are related by the
\emph{stoichiometric matrix} $N$:
\begin{align}
 \dX &= N V\label{eq:dX}
\end{align}
In the case of figure \ref{fig:simple-ic}(a):
\begin{xalignat}{3}
 X &=
 \begin{pmatrix}
   x_a\\x_b
 \end{pmatrix}&
 V &= v&
 \text{and }
 N =
 \begin{pmatrix}
   -1\\1
 \end{pmatrix}  \label{eq:N_a}
\end{xalignat}


The system of figure \ref{fig:simple-ic}(b) has 3 \C components and
the state $X$ can be chosen as:
\begin{equation}
 \label{eq:X_abc_i}
  X =
 \begin{pmatrix}
   x_a&x_b&x_c
 \end{pmatrix}^T
\end{equation}
There are two reaction flows $v_1$ and $v_2$ corresponding to \BRe{1}
and \BRe{2} respectively. The flow vector $V$
can be chosen as:
\begin{equation}
 \label{eq:V_abc_i}
 V =
 \begin{pmatrix}
   v_1&v_2
 \end{pmatrix}^T
\end{equation}
Following the causal strokes and observing the sign convention at the
\Zero~junction:
\begin{xalignat}{3}
 \dx_a &= v_a = -v_1&
 \dx_b &= v_b =  v_1 - v_2&
 \dx_c &= v_c =  v_2
\end{xalignat}
Using (\ref{eq:X_abc_i}) and (\ref{eq:V_abc_i}), it follows that:
\begin{xalignat}{2}
 \dX &= N V \notag&
 \text{where }
 N &=
 \begin{pmatrix}
   -1&0\\
   1&-1\\
   0&1
 \end{pmatrix}\label{N_ABC}
\end{xalignat}
The system of figure \ref{fig:simple-ic}(d) is the same as that of
figure \ref{fig:simple-ic}(b) but with an additional input
$v_0$ and so $V$ is defined as:
\begin{equation}
 \label{eq:V_abcv_i}
 V =
 \begin{pmatrix}
   v_0&v_1 &v_2
 \end{pmatrix}^T
\end{equation}
Using the summing rules at the left and right \Zero~junctions,
it follows that:
\begin{xalignat}{2}
 \dX &= N V \notag&
 \text{where }
 N &=
 \begin{pmatrix}
   1&-1&0\\
   0&1&-1\\
   -1&0&1
 \end{pmatrix}\label{N_ABCv}
\end{xalignat}

The system of figure \ref{fig:simple-ic}(c) has 4 \C components and
the state $X$ can be chosen as:
\begin{equation}
 \label{eq:X_abcd_i}
  X =
 \begin{pmatrix}
   x_a&x_b&x_c&x_d
 \end{pmatrix}^T
\end{equation}
There is one reaction flows $v$ corresponding to \Re. The flow vector
$V$  is thus scalar in this case:
\begin{equation}
 \label{eq:V_abcd_i}
 V = v
\end{equation}
Following the causal strokes and observing the sign convention at the
\Zero~junction:
\begin{xalignat}{4}
 \dx_a &= v_a = -v&
 \dx_b &= v_b = -v&
 \dx_c &= v_c =  v&
 \dx_d &= v_d =  v
\end{xalignat}
Using (\ref{eq:X_abc_i}) and (\ref{eq:V_abc_i}), it follows that:
\begin{align}
 \dX &= N V \notag\\
 \text{where }
 N &=
 \begin{pmatrix}
   -1&-1&1&1
 \end{pmatrix}^T\label{N_ABCD}
\end{align}

In bond graph terms this particular arrangement of causal strokes is
known as \emph{integral causality}. Naturally this analysis extends to
arbitrarily large systems, and can be carried out algorithmically in
automated software.

\subsection{Stoichiometric null spaces}
\label{sec:deriv}
\begin{figure}[htbp]
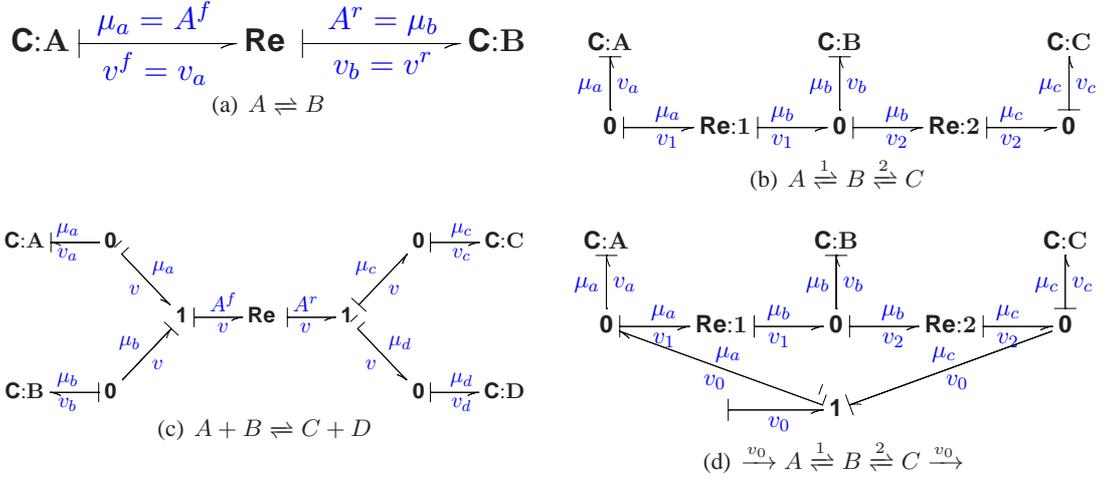

 \centering
 
 \subfigure[$A \reac B$]{
   \resizebox{0.45\linewidth}{!}{\input{AB_dcbg}}
   \label{subfig:AB_dcbg}
 }

 \subfigure[$A \reacu{1} B \reacu{2} C$]{
   \resizebox{0.45\linewidth}{!}{\input{ABC_dcbg}}
   \label{subfig:ABC_dcbg}
 }

 \subfigure[$A+B \reac C+D $]{
   \resizebox{0.45\linewidth}{!}{\input{ABCD_dcbg}}
   \label{subfig:ABCD_dcbg}
 }

 \subfigure[$ \xrightarrow{v_0} A \reacu{1} B\reacu{2} C\xrightarrow{v_0}$]{
   \resizebox{0.45\linewidth}{!}{\input{ABCv_dcbg}}
   \label{subfig:ABCv_dcbg}
 }

  \caption[Causal strokes and the Stoichiometric Matrix Subspaces]
 {Causal strokes and the Stoichiometric Matrix Subspaces. The bond graph notion
   of causality provides an algorithm for determining the
   subspaces of the stoichiometric matrix by explicitly showing how
   the \C flows propagate to other \C flows and
   and to the \Re flows.
   (a) \BC{A}  imposes a flow into \Re which in turn imposes
   a flow into \BC{B}.
   (b) \BC{A} imposes a flow into \BRe{1} and thence, together with
   \BC{B}. imposes a flow into the \Zero~junction and thence into
   \BRe{2} and \BC{C}.
   (c) \BC{A} imposes a flow into the \One~junction and thence into
   \BC{B} and \Re; \Re in turn imposes a flow into the \One~junction
   and thence into \BC{C} and \BC{D}.
   (d) As the external flow $v_0$ impinges on to \Zero~junctions, it
   does not affect the causality of the parts in common with (b).
}
 \label{fig:simple-dc}
\end{figure}
The causal analysis of \S\ref{sec:integral} asks the question:
``given the reaction flows $V$, what are the flows $\dX$ at the \C
components?''. This section looks at the inverse question: ``given the
flows $\dX$ at the \C components, what are the reaction flows $V$?''

With this in mind, the causal stroke on the bond impinging on the
\BC{A} component in figure \ref{fig:simple-dc}(a) is now at the
\C end of the bond, thus imposing flow on the \Re component and so
$v^f=v_a$. There is now a causal issue: as $v_r=v_f$, it follows that
$v_r$ is also determined by the \BC{A} component and $v_r =
v_a$. Hence the flow on the bond impinging on \BC{B} is determined and
the causality must be as shown. Thus causal considerations show that
the flow $v_a$ determines the flow $v_b$ which therefore cannot be
independently chosen. In bond graph terms this particular arrangement
of causal strokes is known as \emph{derivative causality}.
To summarise:
\begin{xalignat}{2}
 v &= v_a = -\dx_a\label{eq:dc_a_v}&
 \dx_b &= v = -\dx_a 
\end{xalignat}

The system of figure \ref{fig:simple-dc}(a) has 2 \C components and
\begin{equation}
 \label{eq:X_ab_d}
  X =
 \begin{pmatrix}
   x_a&x_b
 \end{pmatrix}^T
\end{equation}
It is convenient to decompose $X$ into two components: $x$ the
independent part of $X$ and $X^d$ the
dependent part of $X$. In particular:
\begin{align}
 x &= x_a = L_{xX}X \label{eq:L_xX}
 \text{ where } 
 L_{xX} =
 \begin{pmatrix}
   1&0
 \end{pmatrix}\\
 \text{ and }
 X^d &=
 \begin{pmatrix}
   x_b
 \end{pmatrix}
 = L_{dX}X \label{eq:L_dX}
 \text{ where } 
 L_{dX} =
 \begin{pmatrix}
   0&1
 \end{pmatrix}
\end{align}
The full state $X$ can be reconstructed from $x$ and $X^d$ using:
\begin{equation}\label{eq:X_recon}
 X = L_{xX}^T x + L_{dX}^T X^d
\end{equation}

Using this decomposition Equations
(\ref{eq:dc_a_v}) can be written as:
\begin{align}
 \dX^d &= L_{dx}\dx \label{eq:dc_a_dx}\\ 
 \text{where } 
  L_{dx} &= 
 \begin{pmatrix}
   -1
 \end{pmatrix}
\end{align}

Combining these equations,
\begin{equation}
 \dX^d =  L_{dX}\dX = L_{dx}\dx = L_{dx}L_{xX}\dX
\end{equation}
Defining
\begin{equation}
 \label{eq:G}
 G = L_{dX} - L_{dx}L_{xX}
\end{equation}
it follows that
the state dependency can also be expressed as:
\begin{equation}
 \label{eq:GX_dc_a}
 G\dX = GNV = 0
\end{equation}
where, in this case:
\begin{align}
 G &= 
 \begin{pmatrix}
   0&1
 \end{pmatrix}
 +\begin{pmatrix}
   1&0
 \end{pmatrix}
 = 
\begin{pmatrix}
   1&1
 \end{pmatrix}
\end{align}
As discussed in the textbooks, as (\ref{eq:GX_dc_a}) is true for all
$V$, 
\begin{equation}
 \label{eq:GN}
 GN = 0
\end{equation}
and thus $G$ is a left null matrix of $N$.
In this particular case $G\dX=0$ corresponds to:
\begin{align}
 \dx_a +\dx_b &= 0  \label{eq:GdX_ab}\\
 \text{or }
 x_a + x_b &= \text{const}\notag
\end{align}
Thus the total amount of $A$ and $B$ is constant.

The system of figure \ref{fig:simple-dc}(b) has 3 \C components and
\begin{equation}
 \label{eq:X_abc_d}
  X =
 \begin{pmatrix}
   x_a&x_b&x_c
 \end{pmatrix}^T
\end{equation}
Following the same arguments as for figure \ref{fig:simple-dc}(a), it
follows that:
\begin{align}
 x &= 
\begin{pmatrix}
   x_a\\x_b
 \end{pmatrix}
 = L_{xX}X
 \text{where } 
 L_{xX} =
 \begin{pmatrix}
   1&0&0\\
   0&1&0
 \end{pmatrix}\\
 \text{and }
 X^d &=
 \begin{pmatrix}
   x_c
 \end{pmatrix}
 = L_{dX}X
 \text{ where } 
 L_{dX} =
 \begin{pmatrix}
   0&0&1
 \end{pmatrix}
\text{ and }
  L_{dx} = 
 \begin{pmatrix}
   -1&-1
 \end{pmatrix}
\end{align}
In this case:
\begin{align}
 G &=  L_{dX} - L_{dx}L_{xX}
 =
 \begin{pmatrix}
   0&0&1
 \end{pmatrix}
 -\begin{pmatrix}
   -1&-1
 \end{pmatrix}
 \begin{pmatrix}
   1&0&0\\
   0&1&0
 \end{pmatrix}
 = 
\begin{pmatrix}
   1 & 1 & 1
 \end{pmatrix}
\end{align}
In this particular case $G\dX=0$ corresponds to:
\begin{align}
 \dx_a + \dx_b + \dx_c &= 0  \label{eq:GdX_abc}\\
 \text{or }
 x_a + x_b + x_c &= \text{const}\notag
\end{align}
Thus the total amount of $A$, $B$ and $C$ is constant.

The system of figure \ref{fig:simple-dc}(c) has 4 \C components and
\begin{equation}
 \label{eq:X_abcd_d}
 X =
 \begin{pmatrix}
   x_a&x_b&x_c&x_d
 \end{pmatrix}^T
\end{equation}
Following the same arguments as for figure \ref{fig:simple-dc}(a), it
follows that:
\begin{align}
 x &= 
\begin{pmatrix}
   x_a
 \end{pmatrix}
 = L_{xX}X
 \text{ where } 
 L_{xX} =
 \begin{pmatrix}
   1&0&0&0
 \end{pmatrix}\\
 \text{and }
 X^d &=
 \begin{pmatrix}
   x_b&x_c&x_d
 \end{pmatrix}^T
 = L_{dX}X\\
 \text{where } 
 L_{dX} &=
 \begin{pmatrix}
   0&1&0&0\\
   0&0&1&0\\
   0&0&0&1\
 \end{pmatrix}
\text{ and }
  L_{dx} = 
 \begin{pmatrix}
   1\\-1\\-1
 \end{pmatrix}
\end{align}
In this case:
\begin{align}
 G &=  L_{dX} - L_{dx}L_{xX}
 =
\begin{pmatrix}
   0&1&0&0\\
   0&0&1&0\\
   0&0&0&1\
 \end{pmatrix}
 -  \begin{pmatrix}
   1\\-1\\-1
 \end{pmatrix}
   \begin{pmatrix}
   1&0&0&0
 \end{pmatrix}
 = 
\begin{pmatrix}
  -1&1&0&0\\
  1&0&1&0\\
  1&0&0&1
\end{pmatrix}
\end{align}
In this particular case, $G\dX=0$ corresponds to:
\begin{xalignat}{3}
  x_b  &= x_a + \text{const}\notag&
  x_a + x_c &= \text{const}\notag&
  x_a + x_d &= \text{const}\notag
\end{xalignat}
Thus the amount of $B$ equals the amount of $A$ plus a constant, the
total amount of $A$ and $C$ is constant and the total amount of $A$
and $D$ is constant.

Continuing the analysis of the system of figure \ref{fig:simple-dc}(b) but
including the extra input of figure \ref{fig:simple-dc}(d), the flow
vector has an extra component $v_0$ and can be defined as: 
\begin{equation}
 V = 
 \begin{pmatrix}
   v_0&v_1&v_2
 \end{pmatrix}^T
\end{equation}
where $v_1$ and $v_2$ are the two reaction flows. It is convenient to
decompose $V$ into two components: $v$ the independent part of $V$ and
$V^d$  part of $V$ dependent on $\dX$ and $v$.  In particular:
\begin{align}
 v &= v_0 = K_{vV}V \label{eq:K_vV}
 \text{ where } 
 K_{vV} =
 \begin{pmatrix}
   1&0&0
 \end{pmatrix}\\
  V^d &=
 \begin{pmatrix}
   v_1\\v_2
 \end{pmatrix}
 = K_{dV}V \label{eq:K_dV}
 \text{ where } 
 K_{dV} =
 \begin{pmatrix}
   0&1&0\\
   0&0&1
 \end{pmatrix}
\end{align}

Moreover, following the causal strokes in figure \ref{fig:simple-dc}(d)
the flow vector $V$ can be written in terms of the state
derivative $\dX$ and the independent flow $v$
as:
\begin{align}
 V &= K_{VX}\dX + K_{Vv}v\\
 \text{where }
  K_{VX} &=
  \begin{pmatrix}
    0 &  0  & 0\\
    -1 &  0 &  0\\
    -1 & -1 &  0
  \end{pmatrix}
 \text{ and }
  K_{Vv} =
  \begin{pmatrix}
    1\\1\\1
  \end{pmatrix}
\end{align}
In the particular case that the system is in a steady state and so
$\dX=0$:
\begin{align}
 V &= Kv
\text{ where }
K = K_{Vv}
\end{align}
Substituting into Equation (\ref{eq:dX}) it follows that $NKv = 0$.
As this must be true for all $v$, it follows that $NK = 0$
and thus $K$ is a right null matrix of $N$.

\subsection{Reduced-order equations}
\label{sec:reduced}
The stoichiometric analysis of \S\S\ref{sec:integral} and
\ref{sec:deriv} has many uses; one of these, reducing the order of the
ODEs describing a system\footnote{Order reduction is also discussed by
 an number of authors including \citep{Red88,IngSau03,Ing04,Sau09,Ing13}},
is given here. Reducing system order gives a smaller set of equations
to solve and may avoid numerical problems, for example arising from failure to recognise conserved moieties in a reaction system.

From Equation (\ref{eq:dc_a_dx}), the derivatives $\dX^d$ of the
dependent state $X^d$ can be written as linear transformation of the
derivatives $\dx$ of the independent state $x$ as:
\begin{equation}
 \label{eq:dXd}
 \dX^d = L_{dx}\dx 
\end{equation}
Integrating this equation gives:
\begin{equation}
 \label{eq:Xd}
 X^d - X^d(0) =  L_{dx} ( x - x(0) )
\end{equation}
where $X^d(0)$ and $x(0)$ are the values of $X^d$ and $x$ at time
zero.  Using Equations \ref{eq:L_xX}, \ref{eq:L_dX} and (\ref{eq:G}),
Equation (\ref{eq:Xd}) can be rewritten as:
\begin{align}
 X^d  &=  L_{dx} x  + X^d(0) - L_{dx} x(0)
 = L_{dx} x + ( L_{dX} - L_{dx}  L_{xX} ) X(0)
 = L_{dx} x + G X(0)\label{eq:X_d_from_x}
\end{align}
Using Equation (\ref{eq:X_recon}) to reconstruct $X$ from $X^d$ given
by Equation (\ref{eq:X_d_from_x}) and $x$ gives:
\begin{align}
 X &= \left (
   L_{xX}^T +  L_{dX}^T L_{dx}
 \right ) x 
 + L_{dX}^T G X(0)
 = L x + G_XX(0)\label{eq:X_from_x}\\
 \text{where } 
 L &= L_{xX}^T +  L_{dX}^T L_{dx}
 \text{ and } 
 G_X = L_{dX}^T G \label{eq:G_X}
\end{align}
Equation (\ref{eq:X_d_from_x}) gives an explicit expression for
reconstructing the full state $X$ from the independent state $x$ and
the initial state $X(0)$.

From Equation (\ref{eq:dX}) the state $X$ is given by the system ODE as:
\begin{equation}
 \label{eq:ode_X}
 \dX = NV(X,u)
\end{equation}
where $u$ represents external flows (for example $v$ in figure
\ref{fig:simple-dc} (d)).
Using Equations (\ref{eq:L_xX}) and (\ref{eq:X_from_x}), the ODE
in $X$ of Equation (\ref{eq:ode_X}) can be rewritten as the reduced
order ODE in $x$ as:
\begin{equation}
 \label{eq:ode_x}
 \dx = L_{xX}N V(L x + G_XX(0),u)
\end{equation}
and the full state reconstructed using Equation (\ref{eq:X_from_x}).

\section{Model Reduction and Approximation of Reaction Mechanisms}
\label{sec:approximation}
As discussed in the introduction, complex systems can be simplified by
approximation. However, it is crucial that such approximation does not
destroy the compliance with thermodynamic principles reflected in the
original system.


\begin{figure}[htbp]
 \centering
 \SubFig{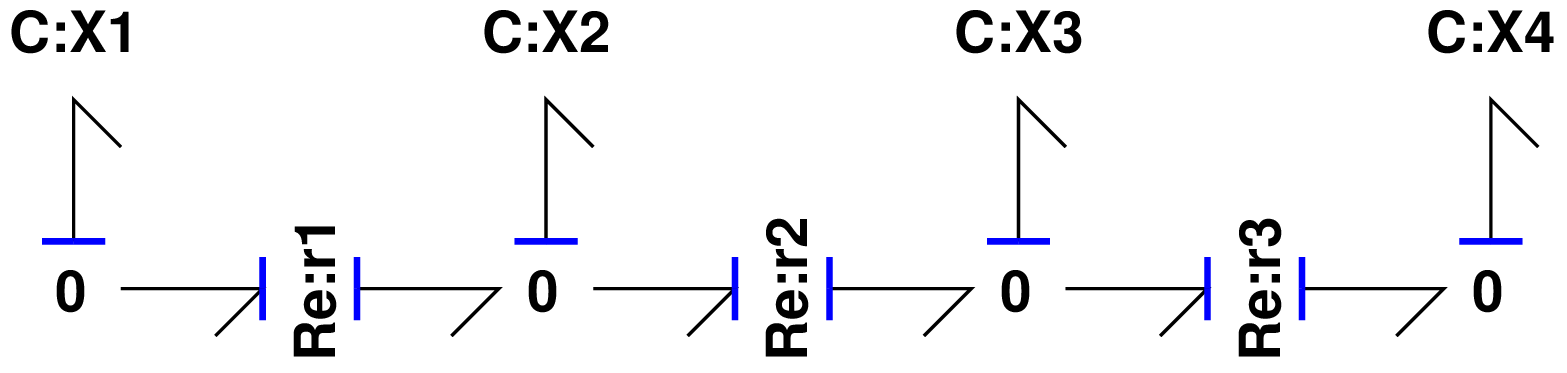}{Full system}{0.45}
 \SubFig{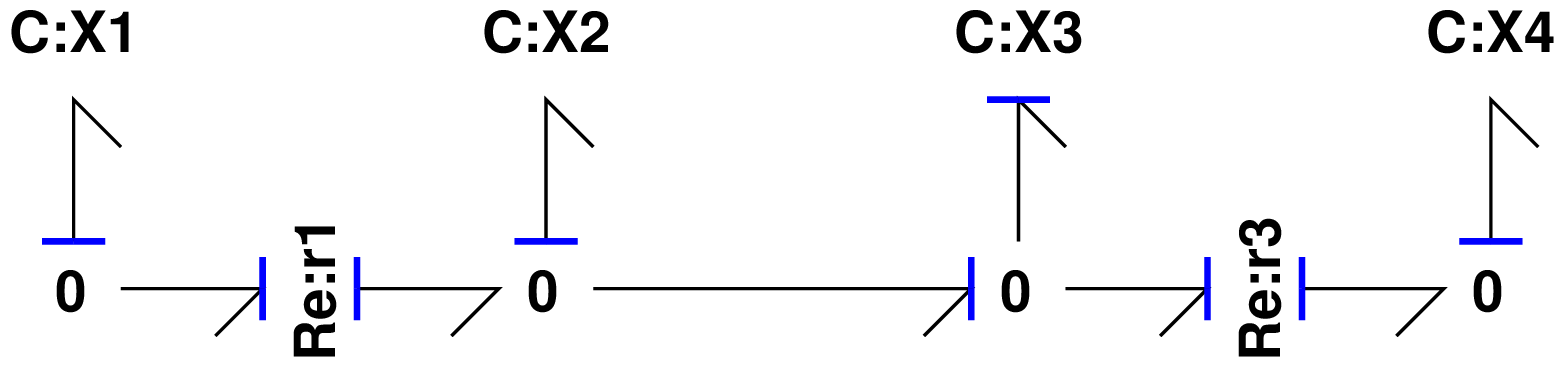}{Approximate system}{0.45}
 \SubFig{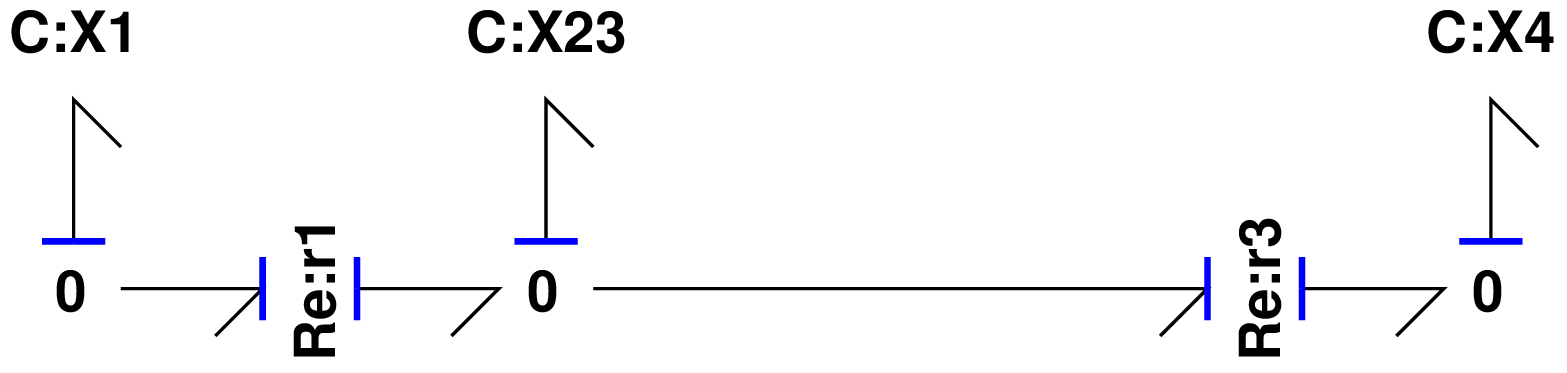}{Equivalent approximate system}{0.45}
 \caption[Approximation of a unimolecular reactions]{Approximation of
   unimolecular reactions \citep[\S3.1]{SmiCra04}. The reaction
   chain (a) is approximated in (b) by assuming that the reaction
   represented by \BRe{r2} is fast ($1/\kappa_2 \approx 0$) and so may
   be removed. (c) is exactly equivalent to (b) except that the
   adjacent \BC{X2} and \BC{X3} are replaced by the composite
   component \BC{X23} with coefficient $K_{23}$ given by Equation
   (\ref{eq:K_23})}
 \label{fig:approx_chain}
\end{figure}

In their analysis of the Sodium Pump, which transports sodium ions out of electrically excitable cells such as cardiomyocytes, \citet{SmiCra04} consider simplification of the linear chain of reactions:
\begin{equation}
 \label{eq:chain}
 \cdots X_1 
 \reacul{k_1^+}{k_1^-} X_2 
 \reacul{k_2^+}{k_2^-} X_3
 \reacul{k_3^+}{k_3^-} X_4  
 \cdots
\end{equation}
where the middle reaction in the chain is fast relative to the other reactions. The three reactions have flows $v_1 \dots v_3$ given by:
\begin{xalignat}{3}
 v_1 &= k_1^+ X_1 - k_1^- X_2&
 v_2 &= k_2^+ X_2 - k_2^- X_3&
 v_3 &= k_3^+ X_3 - k_3^- X_4 \label{eq:v_1}
\end{xalignat}
Reaction (\ref{eq:chain}) corresponds to the bond graph of figure
\ref{subfig:Approximation_cbg} which has the flows of Equations
(\ref{eq:v_1}) where:

\begin{xalignat}{2}
 k_1^+ &= \kappa_1 K_1 & k_1^- &= \kappa_1 K_2\\
 k_2^+ &= \kappa_2 K_2 & k_2^- &= \kappa_2 K_3  \label{eq:k_2}\\
 k_3^+ &= \kappa_3 K_3 & k_3^- &= \kappa_3 K_4
\end{xalignat}

If $\kappa_2 \gg \kappa_1$ and $\kappa_2 \gg \kappa_3$ Equation
(\ref{eq:k_2}) can be rewritten as:
\begin{equation}
 \label{eq:fast}
 \kappa_2 = \frac{1}{\epsilon} 
\end{equation}
where $\epsilon$ is a small positive number. $v_2$ (\ref{eq:v_1})
and (\ref{eq:k_2}) can then be rewritten as:
\begin{equation}
 \epsilon v_2   = K_2 X_2 - K_3 X_3 
\end{equation}
assuming non-zero $v_2$ this means that as $\epsilon \rightarrow 0$,
$X_2$ and $X_3$ are in equilibrium and:
\begin{align}
 X_3 &= \rho X_2 
 \text{ where } \rho = \frac{K_2}{K_3} \label{approx_x_3}
\end{align}
This also means that the difference in affinities associated with
reaction 2 is zero:
\begin{equation}
 A^f_2 - A^r_2 = K_2 X_2 - K_3 X_3 = 0
\end{equation}
Thus the corresponding reaction component \BRe{r3} can be removed
from the bond graph to give figure \ref{subfig:dApproximation_cbg}.
This implies that the \BC{X3} component is in derivative causality and
thus the bond graph represents a differential-algebraic equation
and an ordinary differential equation. However, as discussed by
\citet{GawBev07}, as \BC{X2} and \BC{X3} are on adjacent \zero
junctions, they may be replaced by the single  \BC{X23} component as in
figure  \ref{subfig:aApproximation_cbg}.

Figure \ref{subfig:aApproximation_cbg} represents the same system as
figure \ref{subfig:dApproximation_cbg} if \BC{X23} contains the same
molar mass as  \BC{X2} and \BC{X3}. Moreover, using Equations
(\ref{approx_x_3})
\begin{equation}
 \label{eq:X_23}
 X_{23} = X_2 + X_3 = (1 + \rho) X_2
\end{equation}
The equilibrium constant $K_{23}$ of  \BC{X23} must also correspond to
those of \BC{X2} and \BC{X3} so that:
\begin{align}
 K_2 X_2 &=  K_3 X_3 = K_{23} X_{23}\\
\text{hence }
K_{23} &= \frac{K_2}{1 + \rho} = \frac{\rho K_3}{1 + \rho}\label{eq:K_23}
\end{align}

The bond graph of figure \ref{subfig:aApproximation_cbg} corresponds
to the reaction scheme \citep[\S3.1]{SmiCra04}:
\begin{equation}
 \label{eq:achain}
 \cdots X_1 
 \overset{\alpha_1^+}{\underset{\alpha_1^-}{\rightleftharpoons}} X_{23}
 \overset{\alpha_3^+}{\underset{\alpha_3^-}{\rightleftharpoons}} X_4
 \cdots
\end{equation}
where
\begin{xalignat}{2}
 \alpha_1^+ &= \kappa_1 K_1 = k_1^+ &
 \alpha_1^- &= \kappa_1 K_{23} 
 = \kappa_1 \frac{K_2}{1 + \rho}
 = \frac{k_1^-}{1 + \rho}\notag\\
 \alpha_3^+ &= \kappa_3 K_{23} 
 = \kappa_1 \frac{\rho K_3}{1 + \rho} 
 = \frac{\rho k_3^+}{1 + \rho}
 = \frac{k_3^+}{1 + \frac{1}{\rho}}&
 \alpha_3^- &= \kappa_3 K_4 = k_3^- \label{eq:alpha_1}
\end{xalignat}
Noting that ``$K_2$'' in \citep[\S3.1]{SmiCra04} corresponds to
``$\rho$'' in this paper, Equations
(\ref{eq:alpha_1}) correspond to Equation (18) of
\citet{SmiCra04}.

In general, a chain of $N$ \C components and $N-1$ \Re components
where all of the reactions are fast may be approximately replaced by a
single \C component with:
\begin{align}
 K &= \frac{1}{\frac{1}{K_1} + \frac{1}{K_2} \dots \frac{1}{K_N}}
 = \frac{1}{\sum_{i=1}^N \frac{1}{K_i}}   \label{eq:k_approx}
\end{align}

This procedure is extended to bimolecular reactions in {\S}B of the electronic supplementary material.

\section{Biochemical Cycles}
\label{sec:cycles}
Many biochemical processes central to cellular physiology represent biochemical cycles: including enzyme catalysed reactions, transport processes and signalling cascades. 
A very simple, but practically important, biochemical cycle is the
enzyme-catalysed reaction of figure \ref{subfig:ECR_bg}
This reaction is closely related to that of figure \ref{subfig:ABCD_bg} with 
the important difference that the enzyme $E$ appears on both sides of the reaction
creating the ``loop'' in the bond graph corresponding to a biochemical
cycle. Moreover, in figure \ref{subfig:ECR_bg}, the net flow in
to $E$ is zero and thus $\dx_e = 0$ and $x_e = e_0$ where $e_0$ is a
constant. It follows that:
\begin{align}
v &= \kappa  \left( 
   K_e x_e K_s x_s - K_e x_e K_p x_p
 \right) 
 =  \kappa_e  \left( 
   K_s x_s - K_p x_p
 \right)  \label{eq:v_ex_e}\\
\text{ where }
\kappa_e &=  \kappa K_e e_0 \label{eq:kappa_e}
\end{align}

\subsection{Example: enzyme-catalysed reaction cycles}
\label{sec:MM}
\begin{figure}[htbp]
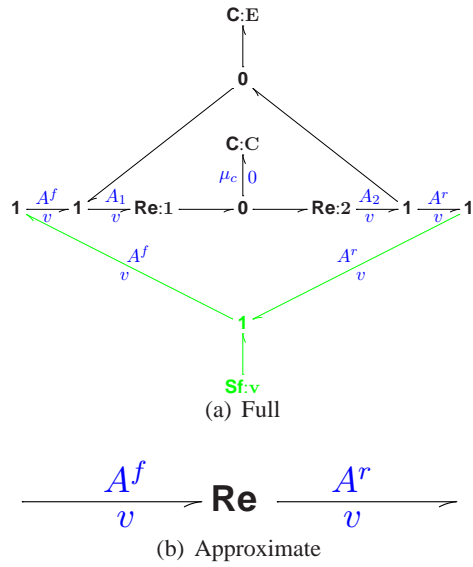

  \centering
  
 \subfigure[Full]{
   \resizebox{0.4\linewidth}{!}{\input{eECR_bg}}
   \label{subfig:eECR_bg}
 }
\\
  
 \subfigure[Approximate]{
   \resizebox{0.4\linewidth}{!}{\input{ECRa_bg}}
   \label{subfig:ECRa_bg}
 }

  \caption{The Michaelis-Menten approximation.}
  \label{fig:MM}
\end{figure}
%
As noted above the enzyme-catalysed reaction of
figure \ref{subfig:ECR_bg} simplifies to a simple reaction with a
modified reaction constant $\kappa_e=\kappa K_e e_0$. However, it is
known from experiments that this simple model of an enzyme-catalysed
reaction fails for high reaction flows. For this reason, as discussed
in the textbooks \citep{KeeSne09,BeaQia10,KliLieWie11}, an
intermediate complex $C$ is introduced so that the reaction:
\begin{equation}
 \label{eq:ECR0}
 S+E \reac P+E
\end{equation}
is replaced by:
\begin{equation}
 \label{eq:ECR}
 S+E \reacu{1} C \reacu{2} P+E
\end{equation}
This reaction may then be replaced by various versions of the
Michaelis-Menten approximation. As discussed by \citet{Gun14} this
approximation has been much misused. In particular, it is used in
circumstances which violate the fundamental law of thermodynamics.

Using the bond graph approach, this section derives a
Michaelis-Menten approximation which is thermodynamically compliant.
In particular, the aim of the approximation is, as for the simple case
of figure \ref{subfig:ECR_bg}, to replace the enzyme-catalysed reaction by a single \Re
component with an equivalent gain $\kappa_e$. But, unlike the simple
case, $\kappa_e$ is not a constant but rather an non-linear function
of the forward and backward affinities.

Figure \ref{fig:MM}(a) shows the enzyme-catalysed reaction (with
complex $C$). The substrate $S$ and product $P$ are omitted from the
bond graph as they do not form part of the approximation. This is a
more general approach than usual as the result to be derived holds for
any biochemical network giving rise to $A^f$ and $A^r$.
As already stated, the aim of the approximation is to replace the bond
graph of figure \ref{fig:MM}(a) by a single \Re component of figure
\ref{fig:MM}(b). As, by definition, the \Re component has the same
flow on each port, it is natural to approximate the bond graph of
figure \ref{fig:MM}(a) by enforcing this constraint at the outset.  To
do this, the flow component \BSf{v} is used to impose a flow $v$ on
each port thus generating the corresponding forward $A^f$ and backward
$A^r$ affinities.


With reference to figure \ref{fig:MM}(a), and using Equation
(\ref{eq:Marcelin}), the equation describing the left-hand \Re
component may be rewritten as:
\begin{align}
 e^\frac{A_1}{RT} &= e^\frac{\mu_c}{RT} + \frac{v}{\kappa_1}
= K_c x_c + \frac{v}{\kappa_1}\\
\text{hence }
A_1 &= RT \ln \left (  K_c x_c + \frac{v}{\kappa_1} \right ),\;
A_2 = RT \ln \left (  K_c x_c - \frac{v}{\kappa_1} \right )
\end{align}
It follows that $A^f$ is given by:
\begin{align}
 A^f &= A_1 - \mu_e
     = A_1 - RT\ln K_e x_e
     = RT \ln \frac{K_c x_c + \frac{v}{\kappa_1}}{K_e x_e}
\end{align}
and, similarly
\begin{align}
 A^r  &= RT \ln \frac{K_c x_c - \frac{v}{\kappa_1}}{K_e x_e}
\end{align}
It is convenient to transform $A^f$ and $A^r$ into $v_o^+$ and $v_o^-$
where:
\begin{align}
 v_0^+ &= e^{\frac{A^f}{RT}},\;
 v_0^- = e^{\frac{A^r}{RT}}  
\end{align}
giving
\begin{xalignat}{2}
 v_o^+ &= \frac{K_c x_c + \frac{v}{\kappa_1}}{K_e x_e}&
 v_o^- &= \frac{K_c x_c - \frac{v}{\kappa_2}}{K_e x_e}\label{eq:ECR_vm}
\end{xalignat}
Subtracting these equations gives:
\begin{align}
 v_o^+ - v_o^- &= \frac{ \frac{1}{\kappa_1} + \frac{1}{\kappa_2} }{K_e
x_e} v\\
\text{ hence }
v = \bar{\kappa}K_e x_e \delta_v
\text{ where }
\bar{\kappa} &= \frac{\kappa_1 \kappa_2}{{\kappa_1+\kappa_2}}
\text{ and } \delta_v = v_o^+ - v_o^-\label{eq:ECR_v0}
\end{align}
Multiplying  Equations  (\ref{eq:ECR_vm}) by
$\kappa_1$ and $\kappa_2$ respectively and adding gives:
\begin{align}
 \kappa_1 v_o^+ + \kappa_2 v_o^- &= (\kappa_1  + \kappa_2)\frac{K_c
   x_c}{K_e x_e} \\
 \text{hence } x_c &= \frac{K_e}{K_c}\sigma_v x_e  
 \text{ where }
 \sigma_v = \frac{\kappa_1 v_o^+ + \kappa_2 v_o^-}{\kappa_1  +
   \kappa_2}
 = \frac{\kappa_1 e^{\frac{A^f}{RT}} + \kappa_2 e^{\frac{A^r}{RT}}}{\kappa_1  +
   \kappa_2} \label{eq:ECR_x_c}
\end{align}
Using the feedback loop implied by $x_e = e_0 - x_c$ and Equation (\ref{eq:ECR_x_c}):
\begin{align}
 x_e &= \frac{e_0}{1 + \frac{K_e}{K_c} \sigma_v}\label{eq:ECR_x_e}
\end{align}
Substituting Equation (\ref{eq:ECR_x_e}) into Equation
(\ref{eq:ECR_v0}) gives:
\begin{align}
 v &= \bar{\kappa} \frac{K_ee_0}{1 + \frac{K_e}{K_c} \sigma_v}
 \delta_v 
 = \bar{\kappa} \frac{K_ce_0}{\frac{K_c}{K_e} +  \sigma_v} \delta_v
\end{align}
There are two special cases of interest $\kappa_1 = \kappa_2$ and
$\kappa_1 \gg \kappa_2$. In these two cases, $ \sigma_v$ is given by:
\begin{equation}
 \label{eq:sigma_v_spec}
 \sigma_v =
 \begin{cases}
   \frac{ v_o^+ + v_o^-}{2} = \frac{e^{\frac{A^f}{RT}} + e^{\frac{A^r}{RT}}}{2} & \kappa_1 = \kappa_2\\
    v_o^+ = e^{\frac{A^f}{RT}} & \kappa_1 \gg \kappa_2
 \end{cases}
\end{equation}
Hence the enzyme-catalysed reaction can be approximated by the \Re component with equivalent
gain $\kappa_e$ given by
\begin{align}
 \kappa_e &= e_0 \frac{\bar{\kappa} K_c}{k_m + \sigma_v} 
 \text{ where } k_m = \frac{K_c}{K_e} \label{eq:kappa_ecr}
\end{align}
In contrast to the expression for the simple case (\ref{eq:kappa_e})
$\kappa_e$ is, via $\sigma_v$ (\ref{eq:sigma_v_spec}), a function of
the affinities $A^f$ and $A^r$. 
The fact that $\sigma_v>0$ ensures that the \Re
component corresponding to  Equation (\ref{eq:kappa_ecr}) is
thermodynamically compliant.

In both Equations (\ref{eq:kappa_e}) and (\ref{eq:kappa_ecr}), the
expression for $\kappa_e$ has a factor $e_0$, the (constant) sum of
$x_e$ and $x_c$. In many biochemical situations, the enzyme $E$ is the
product of another reaction. Although Equation (\ref{eq:kappa_ecr}) is
derived for a constant $e_0$, a further approximation would be to allow
$e_0$ to be time varying $e_0 = x_E$
where $x_E$ is the enzyme concentration from an external reaction.
This leads to the concept of the \emph{modulated} \Re, or \mRe
component of figure \ref{fig:MM}(c).  The additional modulating bond
caries two signals: the effort $\mu_E$ where:
\begin{equation}
 \label{eq:mu_e}
 \mu_E = e^\frac{x_e}{RT}
\end{equation}
and a zero flow.  The zero flow means that the modulating bond does
\emph{not} transmit power.
The \mRe component is used to approximate the system of \S\ref{sec:switch}.


\subsection{Example: a biochemical switch}
\label{sec:switch}
\begin{figure}[htbp]
 \centering
 \SubFig{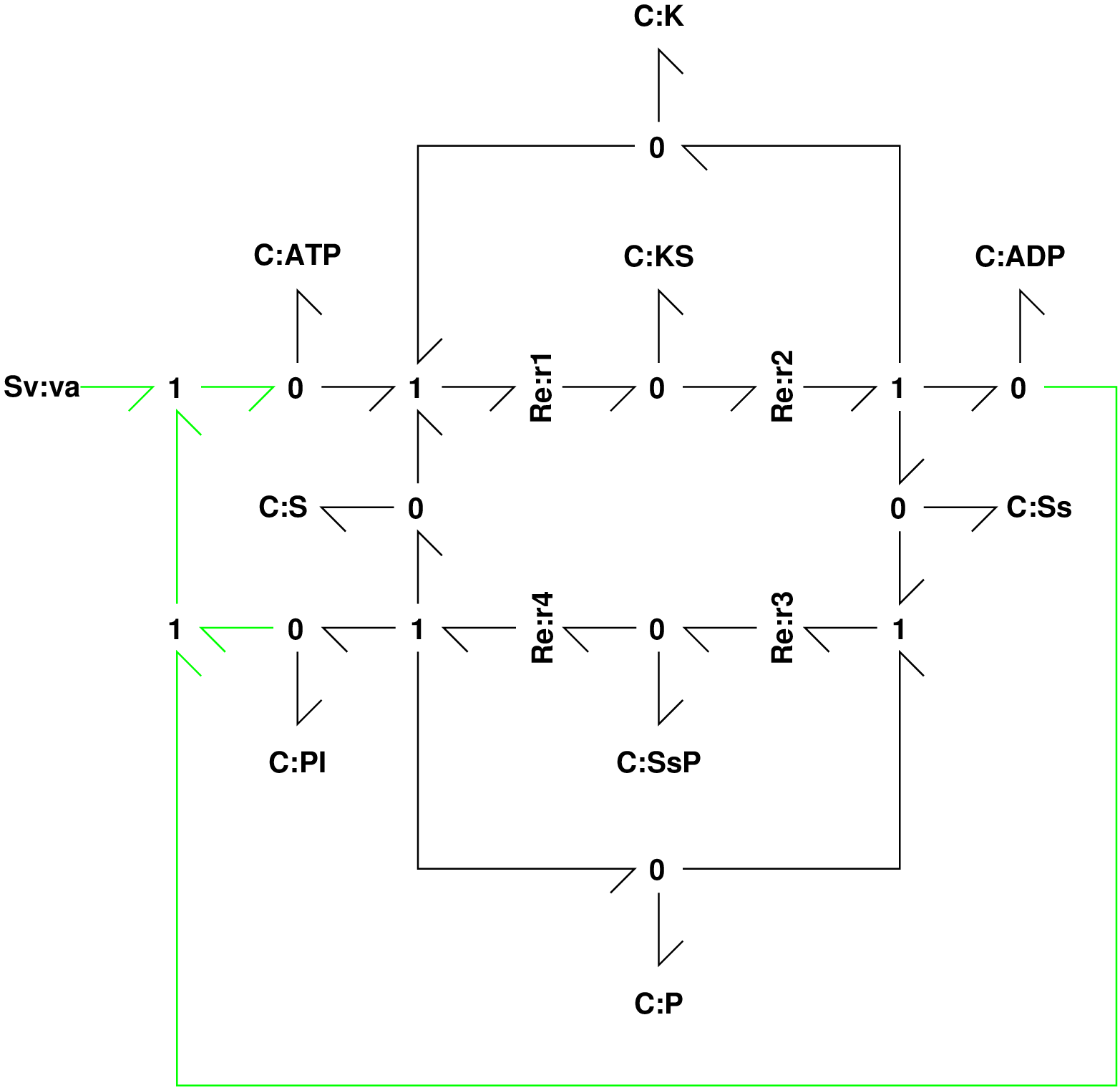}{System bond graph}{0.45}
 \SubFig{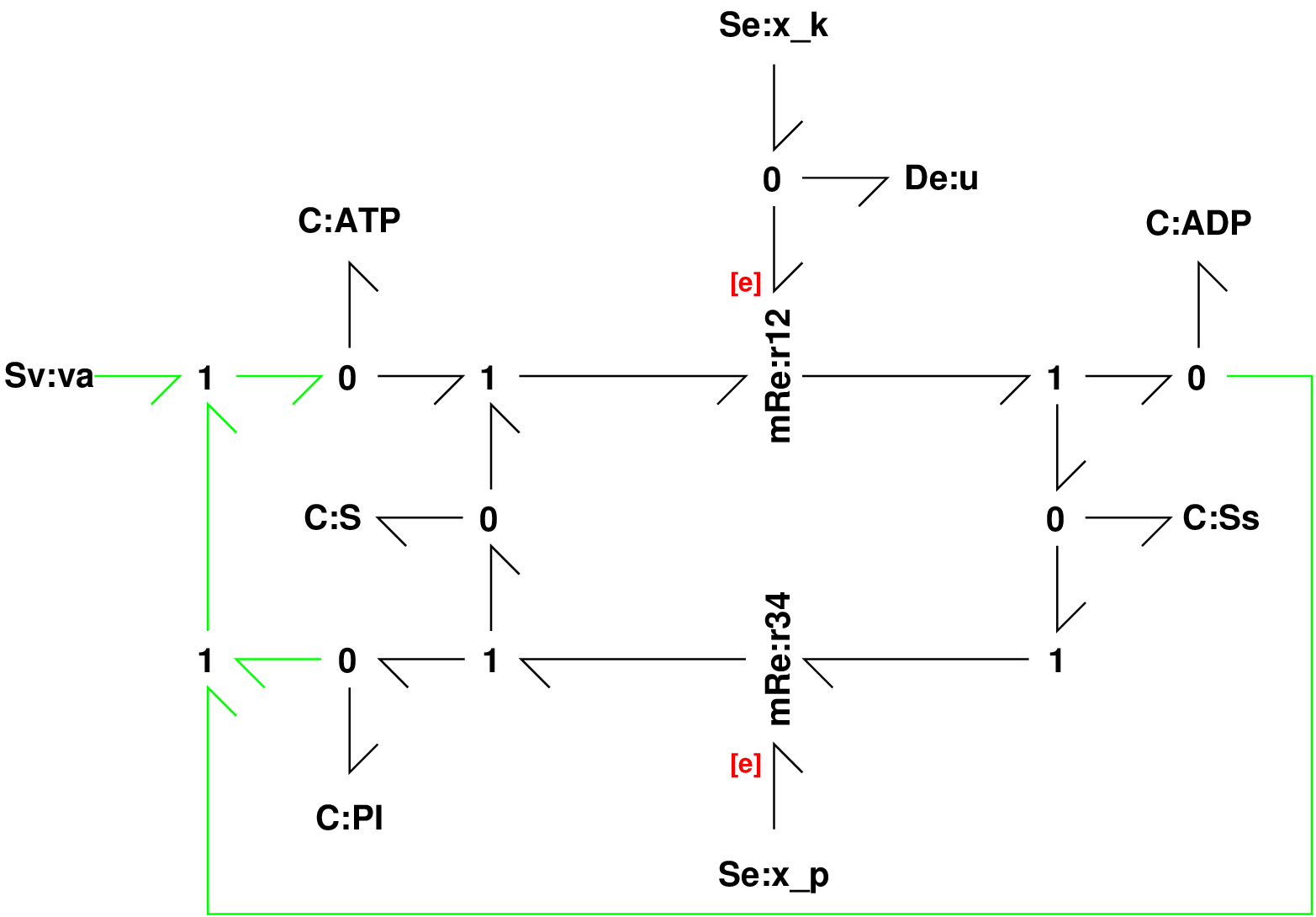}{Approximate system bond graph}{0.45}
 \caption[A Biochemical Switch]{A Biochemical Switch. (a) The bond
   graph of the biochemical switch of \citet{BeaQia10} has four
   reactions \BRe{r1}~--~\BRe{r4} and nine substances. The external
   flow $v_a$ of ATP is required for the long-term operation of the
   switch which consumes ATP. (b) This switch can be approximated
   using the approximation of figure \ref{fig:MM} whist retaining
   thermodynamic compliance.}
 \label{PD_bg}
\end{figure}

\begin{figure}[htbp]
 \centering
 \SubFig{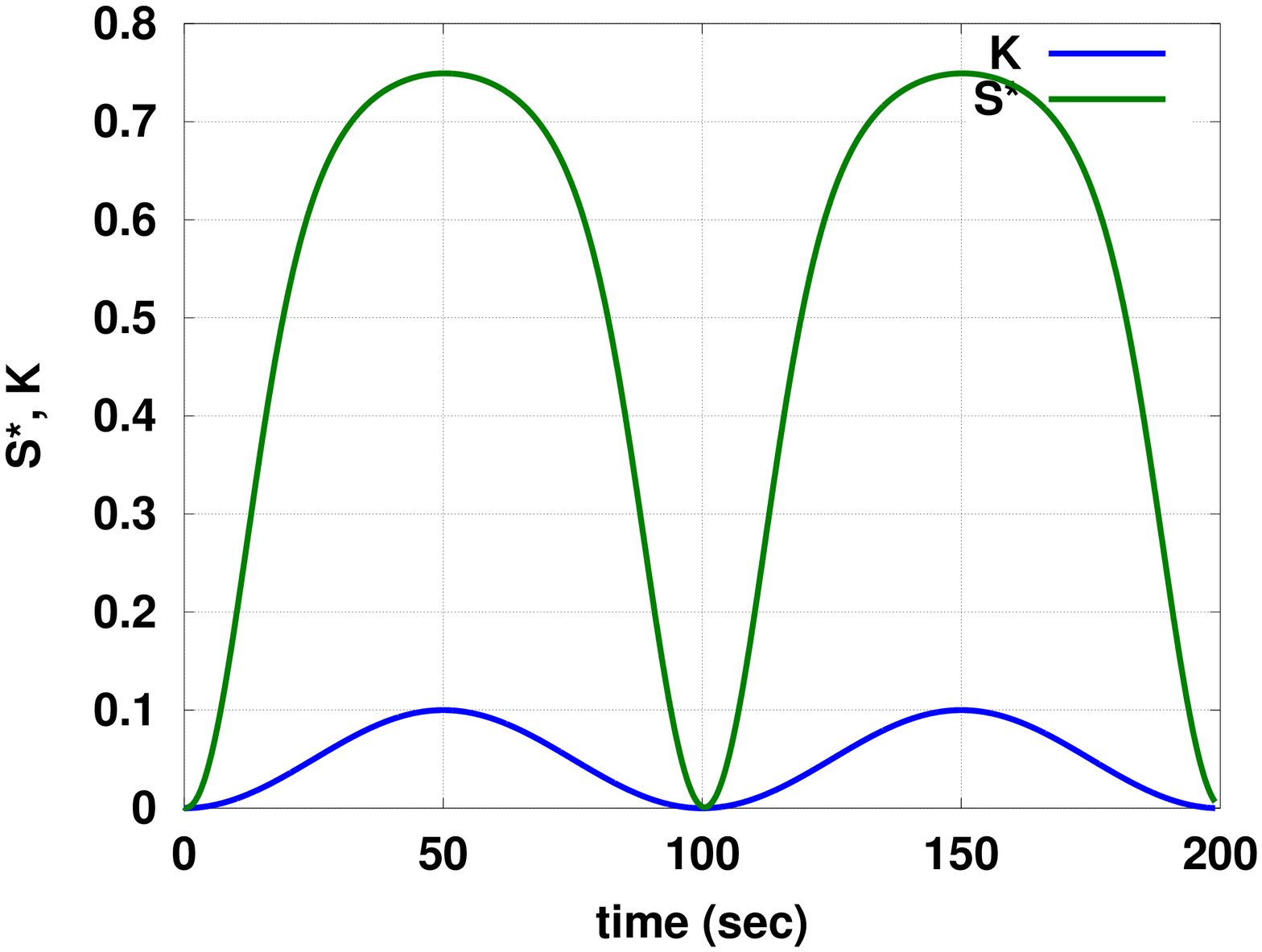}{$S^\star$ \& $K$}{0.45}
 \SubFig{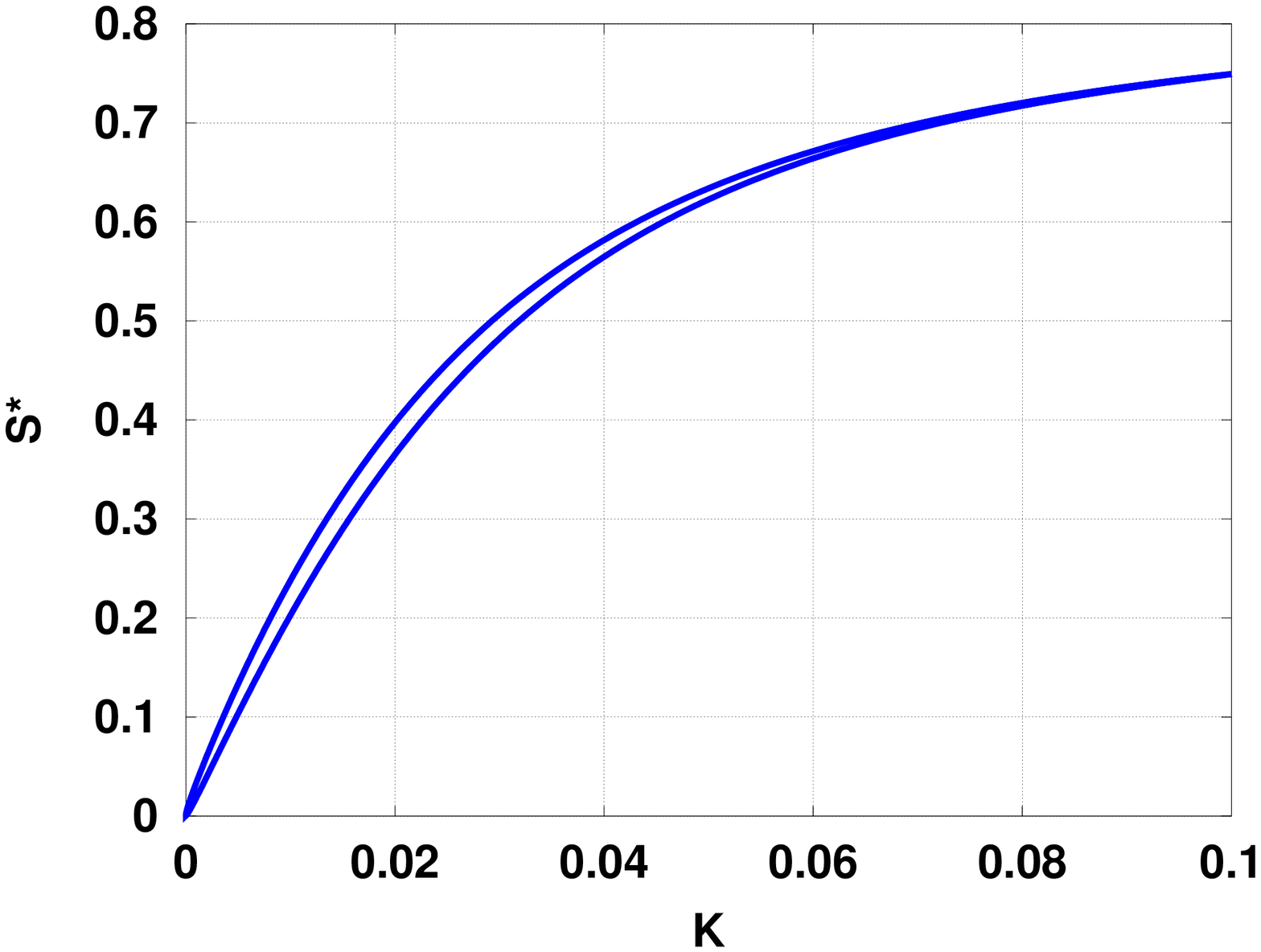}{$S^\star$ v. $K$}{0.45}
 \SubFig{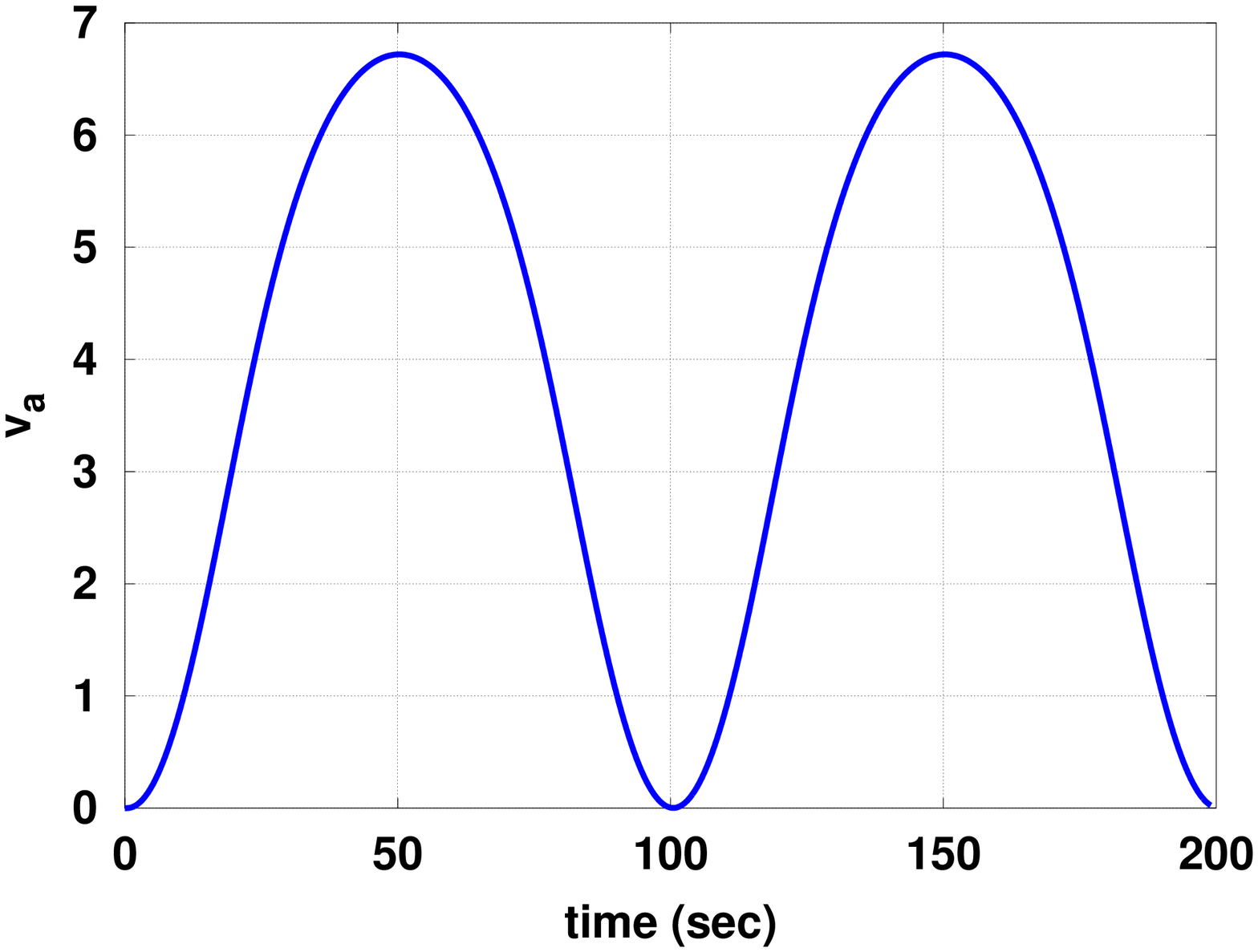}{$v_{a}$ -- ATP flow}{0.45}
 \SubFig{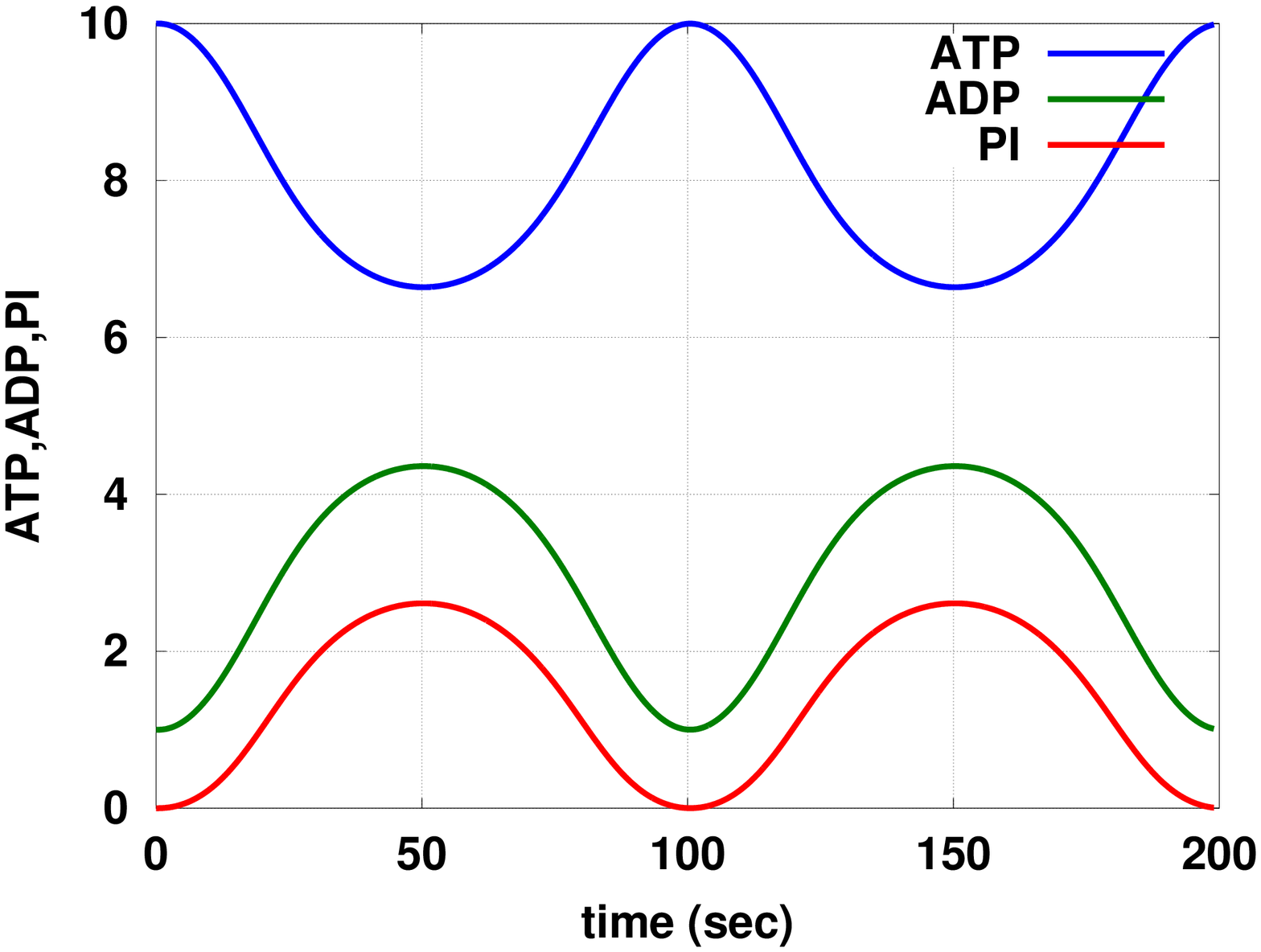}{ATP \& $S$}{0.45}
 \caption[A Biochemical Switch: simulation]{A Biochemical Switch:
   simulation. (a) shows the time response of the amount of $S^\star$
   to a sinusoidal variation in the amount of $K$. The biochemical
   switch both amplifies and distorts the signal. (b) plots $S^\star$
   against $K$ to show the non-linear amplification effect. (c) shows
   the corresponding molar flow of ATP into the system and (d) the
   corresponding amounts of ATP, ADP and Pi.}
 \label{PD_fb}
\end{figure}


\citet[\S5.1.1]{BeaQia10} discuss a biochemical switch described by
\begin{xalignat}{2}
 S + ATP + K &\reacu{1}KS &
 KS &\reacu{2} S^\star + ADP + K \notag\\
 S^\star + P &\reacu{3} S^\star P &
 S^\star P &\reacu{4} S + Pi + P \label{eq:PD_4}
\end{xalignat}
These reactions represent a phosphorylation / dephosphorylation cycle. Protein $S$ is phosphorylated by kinase $K$, and is dephosphorylated by phosphatase $P$, where $S^\star$ represents the phosphorylated (active, perhaps) state of the protein. 
The corresponding bond graph appears in figure \ref{subfig:PD_abg}
where the external flow $v_a$ necessary to top up the ATP reservoir is
included. This system contains 9 states and four reactions with
mass-action kinetics.  Using the approximation of \S\ref{sec:MM}, figure \ref{fig:MM}, this system can be approximated by
the bond graph of \ref{subfig:aPD_abg}. The approximate system has 5
states and two reactions with the reversible Michaelis-Menten kinetics
of \S\ref{sec:MM}. It has the further advantage that the
dynamics are explicitly modulated by the concentrations $x_k$ and
$x_p$ of $K$ and $P$ respectively.

The bond graph of \ref{subfig:aPD_abg} clearly shows a biochemical
cycle. It's behaviour can be understood as follows. When $x_k$ is large.
ATP drives $S$ though the reaction component \BRe{r12} to create
$S^\star$; and this flow is greater than that though \BRe{r34} and so
the amount of $S^\star$ increases at the expense of $S$. However, when
$x_k$  is small the flow though \BRe{r12} becomes less than that
though \BRe{r34} and amount of $S^\star$ decreases.

For the purposes of illustration, the following parameter values were
used. With reference to Equation (\ref{eq:kappa_ecr}), $k_m=0$ and
$\bar{\kappa} K_c=100$ for both reactions. With reference to Equations
(\ref{eq:mu_a}), $K_{ATP}=10$ and
$K_S=K_{S*}=K_{ADP}=K_{PI}=1$. The initial states were:
$x_{ATP}=10$, $x_{S}=x_{ADP}=1$ and $x_{S*}=x_{PI}=0$.

Figure \ref{PD_fb} shows a simulation of the biochemical switch when
ATP is replenished by setting
\begin{equation}
 \label{eq:fb}
 v_a = g_a \left ( w_{atp} - x_{atp} \right )
 \text{ where } g_a = 2
 \text{ and } w_{atp} = 10 
\end{equation}
Equation (\ref{eq:fb}) represents simple proportional feedback;
\emph{in vivo}, this would  correspond to a cellular control system.
Figure \ref{subfig:aPD_fb_tS} shows the response of the amount of
$S^\star$ to a sinusoidal variation in the amount of $K$. The
biochemical switch both amplifies and distorts the signal. This effect
is further shown in figure \ref{subfig:aPD_fb_uS} where the amount of
$S^\star$ is plotted against the amount of $K$. This is basically a
high-gain  saturating function. The hysteresis is due to the time
constant of the feedback loop implied by Equation (\ref{eq:fb}); the
hysteresis reduces if either $g_a$ is increases or the frequency of
the input sinusoid decreased. 
All biochemical cycles require free-energy transduction
\citep{Hil89}. Figure \ref{subfig:aPD_fb_va} shows the molar flow of
ATP into the system (and, as indicated in figure
\ref{subfig:aPD_abg} the outflow of ADP and Pi) as a function of
time; the ON state of the switch induces a flow of ATP using
Equation (\ref{eq:fb}) to replenish the ATP consumed by the cycle.
Figure \ref{subfig:aPD_fb_ATP} shows the corresponding amounts of
ATP, ADP and Pi. The controller does not exactly hold ATP at
the desired level of $w_{atp} = 10$; a higher gain controller would
reduce the control error.
As discussed by \citet[\S5.1.1]{BeaQia10}: ``$\dots$ a
biochemical switch cannot function without a free energy input. No
energy, no switch''. This can be simulated by setting $g_a=0$ in
Equation (\ref{eq:fb}) and forms Figure 1 of \S{A} of the electronic supplementary material.

Approximate models of signalling network components have been
advocated by \citet{KraSolSau10} and \citet{RyaHolDel12} as an
approach to understanding the behaviour of complex signalling
networks. The models developed in this section could also be used for
such a purpose, but with the advantage that the resulting model is
thermodynamically compliant.

Model reduction of an enzymatic cycle model of the SERCA pump
\citep{{TraSmiLoiCra09}} is discussed in {\S}C of the electronic supplementary material.

\section{Hierarchical Modelling of Large Systems}
\label{sec:hierarchical}

One of the objectives of systems biology is to represent the network
of biochemical reactions taking place in cells by computational
models. Large-scale models of cellular metabolic and signalling
networks have been constructed; for example, cardiac cell models which
integrate electrophysiology, metabolism, signalling, and cellular
mechanics have been developed in order to study cell physiology in
normal and disease conditions \citep{FinNieCra11}.

In order to facilitate the development and reuse of such models, XML-based markup languages such as CellML \citep{Lloyd:2004cv} and SBML \citep{Hucka:2003fs} have been created. These languages enable mathematical descriptions of biological processes to be stored in machine-readable formats, but put relatively little restriction on the formulation of the models themselves. 

For example, CellML, which was originally developed in order to share
models of cardiac cell dynamics, represents models as a number of
component elements, each of which contains a number of variables (for
example representing cell membrane potential, or an ionic
concentration), the mathematical relationship between these variables
(for example, the Nernst potential given as a function of the
concentrations) expressed in MathML, and associated parameters. Such
components can be connected to one another to form a model.

This construction allows a modular approach to modelling in which
cellular processes and reactions can be broken down into components,
which are then connected to form a model of the system under study
\citep{CooHunCra08}.
However, there is no requirement that components adhere to the principles of conservation of mass, conservation of charge, or thermodynamic consistency. Nor is there currently any framework which would ensure thermodynamic consistency, or mass or charge conservation, for a model created by connecting components in this modular fashion, even if the components themselves were constructed as thermodynamic cycles. 

The Bond Graph approach which we have outlined here provides such a
framework for modular representation of components of biological
systems, which can be assembled so as to preserve thermodynamic
properties, charge and mass conservation, both in the individual
components and in the overall system. Furthermore, the development of
the Bond Graph Markup Language (BGML) by \citet{Bor06} for the
exchange and reuse of bond graph models, and associated software,
provides the tools through which integration with representations such
as CellML may be achieved.

The stoichiometric analysis of Section \ref{sec:causality}, and its
relationship to causality, is illustrated by simple systems. However,
the notion of bond graph causality, and the corresponding propagation
of causality using the \emph{sequential causality assignment
  procedure} \citep[Chapter
5]{KarMarRos12}, is applicable to arbitarily large systems.

\section{Conclusion}
\label{sec:conclusion}

Based on the seminal work of \citet{OstPerKat71}, the fundamental
concepts of network thermodynamics have been combined with more recent
developments in the bond graph approach to system modelling to give a
new approach to building dynamical models of biochemical networks
within which compliance with thermodynamic principles is automatically
satisfied.
As noted in the Introduction, the bond graph is more than a sketch of
a biochemical network; it can be directly interpreted by a computer
and, moreover, has a number of features that enable key physical
properties to be derived from the bond graph itself. 
It has been shown that stoichiometric properties, including the
stoichiometric matrix $N$ and the left and right null-space matrices
$G$ and $K$, can be directly derived from the bond graph using the
concept of causality associated with bond graphs. The corresponding
causal paths, when superimposed on the bond graph, directly indicate
both pools (conserved moieties) and steady-state flux paths.
The bond graph methodology includes a framework for approximating
complex systems whilst retaining compliance with thermodynamic
principles and this has been illustrated in two contexts: chains
of reactions and the Michaelis-Menten approximation of
enzyme-catalysed reactions.

As emphasised by \citet{BeaQia10}, living organisms are associated with
\emph{non-equilibrium} steady-states. For this reason, this paper has
emphasised the role of external inputs to biochemical networks
modelled by bond graphs. In particular, the example of
\S\ref{sec:switch}, models a biochemical switch where the role of ATP
as a power source is explicitly integrated into the bond graph model.

The bond graph approach is naturally modular in that networks of
biochemical reactions can be connected by bonds whilst retaining
compliance with thermodynamic principles. Modularity has been
illustrated by simple examples and future work will develop
appropriate software tools to build on this natural modularity.

Biochemical networks have non-linear dynamics which generate phenomena
which cannot be generated by linear systems. Nevertheless, useful
information can be obtained from linear models obtained by
linearisation of non-linear systems. In the
context of engineering systems theory, linearisation has been
considered within the framework of sensitivity theory
\citep{Fra78,RosYus00}.
In the context of biochemical networks, Metabolic Control Analysis
(MCA) \citep{HeiSch96} is based on the sensitivity analysis of
stoichiometric networks. The relationship of MCA to engineering
concepts of sensitivity has been examined by \citet{IngSau03},
\citet{Ing04} and \citet{Sau09}.
\citet{Ing04} has shown that standard engineering sensitivity
theory can be applied to biochemical networks to derive frequency
responses with respect to small perturbations in system parameters.
Sensitivity and linearisation of systems described by bond graphs has
been considered by an number of authors 
\citep{Kar77,Gaw00c,GawRon00b}. The bond graph
approach has the advantage of retaining the system structure.
Future work will look at bond graph based linearisation in the
context of biochemical networks.

This paper has focused on deriving thermodynamically compliant
biochemical reaction networks, and their thermodynamically compliant
approximations, from elementary biochemical equations. It would be
interesting to look at the inverse problem: Is a given ODE model of a
system of biochemical reactions with non mass-action kinetics
thermodynamically compliant and does it have a bond graph
representation?

In addition to stoichiometric analysis, the bond graph approach can be
used to directly investigate structural properties of dynamical
systems such as controllability
\citep{SueDau89,SueDau97} and invertiblity
\citep{NgwScaTho96,NgwGaw99,Gaw00d,MarJar11}. Future work will look at
bond graph based structural analysis in the context of biochemical
networks.

The bond graph approach is based on the notion of power flow. For this
reason, it has been much used for modelling multi-domain engineering
systems with appropriate transducer models to interface domains. Thus
for example: an electric motor or a piezo-electric actuator couples
electrical and mechanical domains and a turbine or pump couples
hydraulic and mechanical domains. We will build on the work of
\citet{LefLefCou99} on chemo-mechanical transduction and the work of
\citet{Kar90} on chemo-electrical transduction to interface
biochemical networks with systems involving muscle and excitable
membranes.

We believe that, when combined with modern software tools, the bond
graph approach provides a significant alternative hierarchical and
modular modelling framework for complex biochemical systems in which
compliance with thermodynamic principles is automatically satisfied.

\section*{Acknowledgements}
Peter Gawthrop would like to thank Mary Rudner for her encouragement
to embark on a new research direction.

This research was in part conducted and funded by the Australian
Research Council Centre of Excellence in Convergent Bio-Nano Science
and Technology (project number CE140100036), and by the Virtual
Physiological Rat Centre for the Study of Physiology and Genomics,
funded through NIH grant P50-GM094503. Peter Gawthrop would like to
thank the Melbourne School of Engineering for its support via a
Professorial Fellowship.

The authors would like to thank the anonymous reviewers for helpful
comments on the manuscript.

\appendix

\section{A biochemical switch : further simulations}
\begin{figure}[htbp]
 \centering
 \SubFig{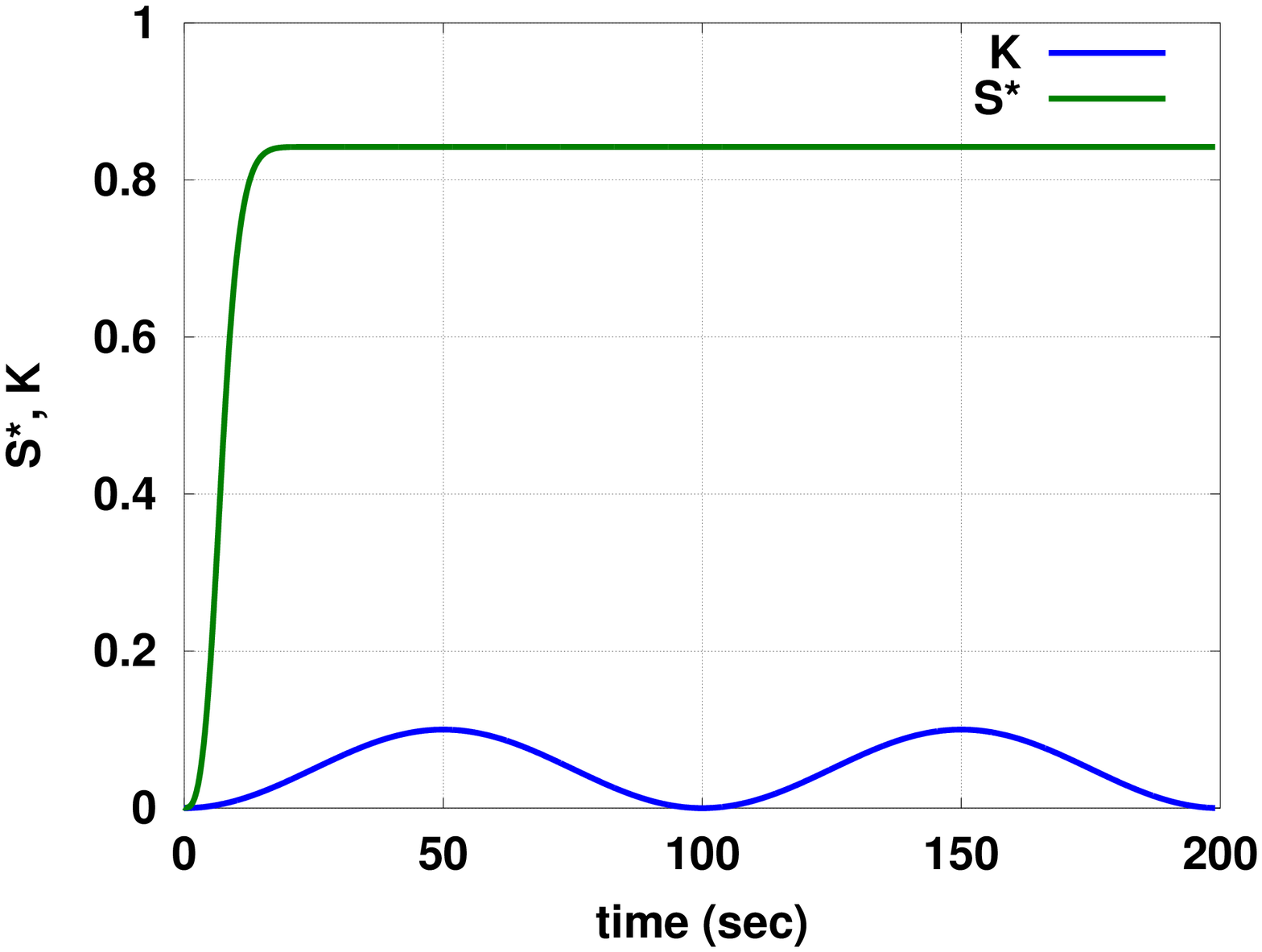}{$S^\star$ \& $K$}{0.45}
 \SubFig{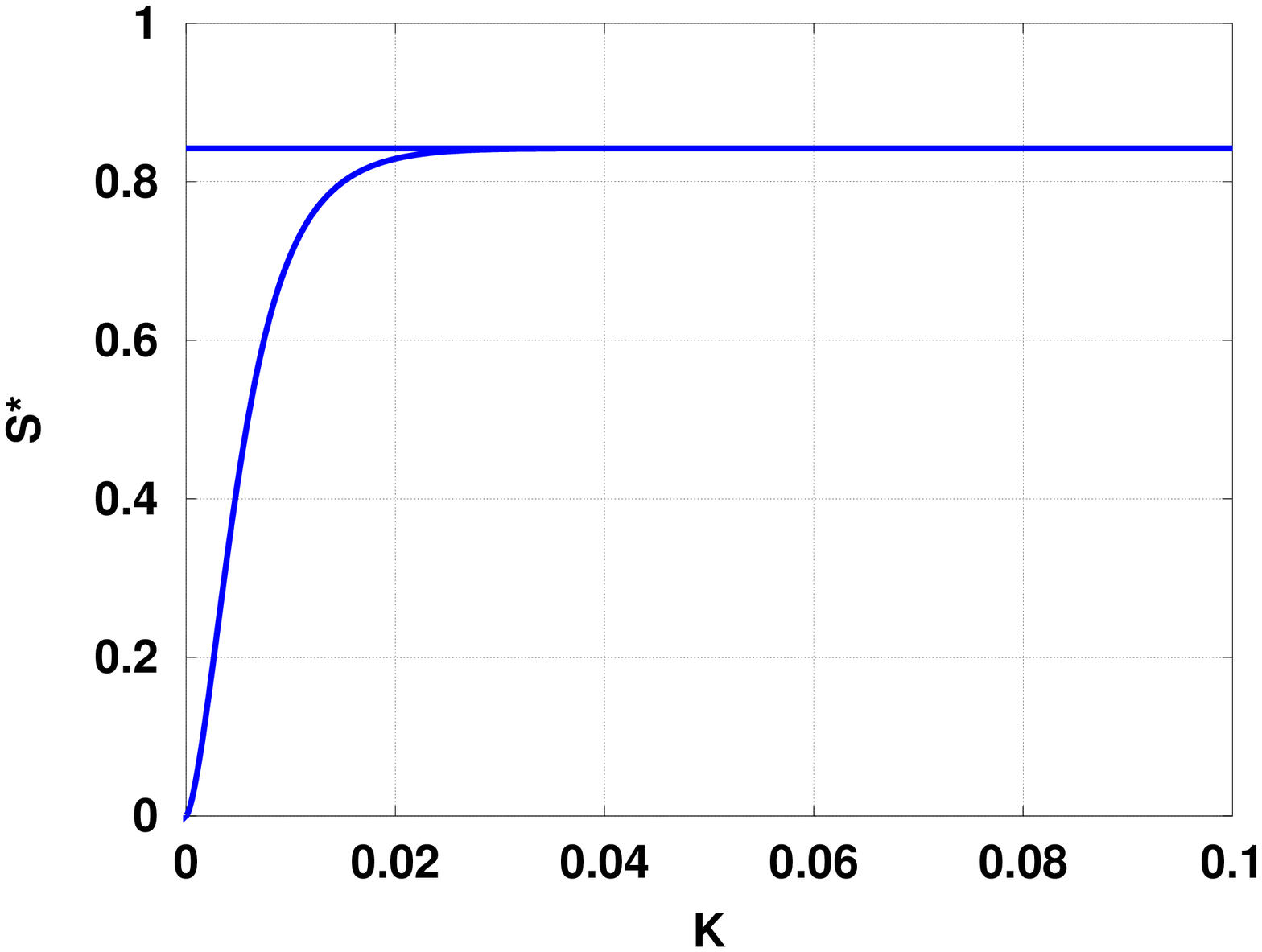}{$S^\star$ v. $K$}{0.45}
 \SubFig{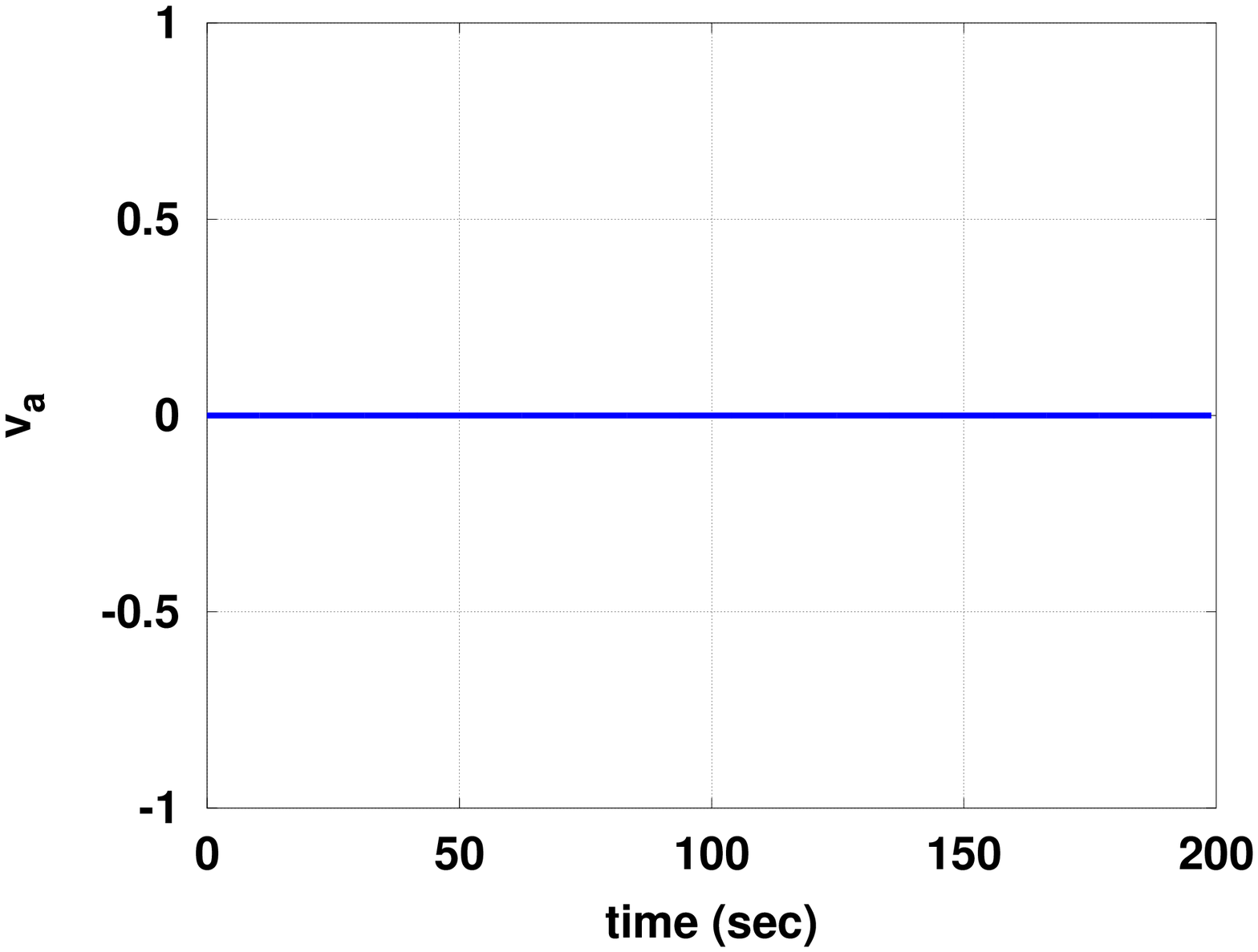}{$v_{a}$ -- ATP flow}{0.45}
 \SubFig{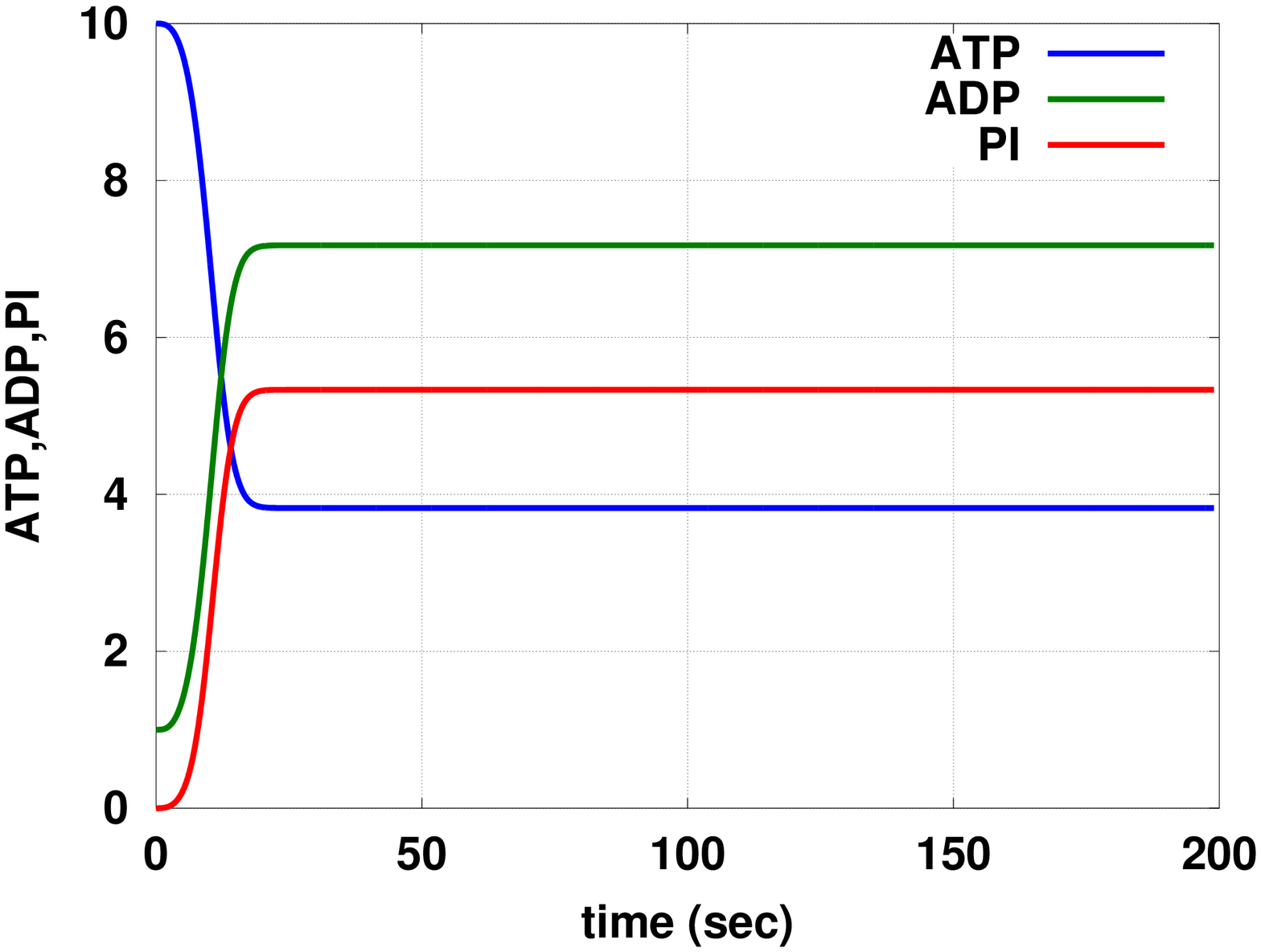}{ATP \& $S$}{0.45}
 \caption[A Biochemical Switch: simulation]{A Biochemical Switch:
   simulation without ATP replenishment. The simulation is the same as
   that of Figure 7 of \S5(b) of the paper except that ATP is not
   replenished and so the switch fails to function.}
 \label{PD_nofb}
\end{figure}

\section{Bimolecular reactions}
\label{sec:bimolecular}
\begin{figure}[htbp]
 \centering
 \begin{tabular}{|m{4.2cm}|m{5cm}|}
   \hline
   Name & Bond Graph\\
   \hline
  (a) Bimolecular reaction & \input{Branch_bg}\\
  (b) Fast bimolecular reaction & \input{Branch_fast_bg}\\
  (c) Simplified bimolecular reaction & \input{Branch_simp_bg}\\
   \hline
 \end{tabular}
 \caption[Approximation of bimolecular reactions.]{Approximation of bimolecular reactions.}
 \label{fig:bimolecular}
\end{figure}
Thermodynamic cycles typically involve bimolecular reactions where, in
bond graph terms, the chain of reactions discussed in Section
4 is augmented by branches. Figure
\ref{fig:bimolecular}(a) shows a reaction of the form $A+C \reac B$
where the presence of $C$ gives the branched bond graph structure. The
bonds at the left and right of Figure \ref{fig:bimolecular}(a) form
connections to the rest of a reaction network.

There are two approximations made to simplify this bimolecular
reaction. As in Section 4 it is assumed that the
reaction is fast and thus the \Re component can be removed as in
Figure \ref{fig:bimolecular}(b). It is further assumed that the
concentration of $C$ is approximately constant either due to
replenishment or to a large pool; this is indicated in Figure
\ref{fig:bimolecular}(b) by replacing the component \BC{C} by \BCS{C}: a
capacitive source. As in Section 4, the removal of the
\Re component changes the causality of  \BC{B}, but the causality of
\BCS{C} remains the same as that of \BC{C}.

Again, the equilibrium induced by the removal of the \Re component
leads to the state $x_b$ of \BC{B} being determined by the states of
the other two components:
\begin{align}
 x_b &= \frac{K_c x_c K_a x_a}{K_b}  \notag\\
 &= \tilde{x}_c x_a   \label{eq:bimolecular:x_b}\\
\text{where } \tilde{x}_c &= \frac{K_c K_a}{K_b}x_c \label{eq:bimolecular:x_c_t}\\
\text{thus } x_{ab} &= x_a + x_b = (1 + \tilde{x}_c) x_a \label{eq:bimolecular:x_ab}
\end{align}

Unlike the unimolecular case, the chemical potentials $\mu_1$ and
$\mu_2$ are different; in particular
\begin{equation}
 \label{eq:mu_2}
 \mu_2 = \mu_1 +  RT \ln K_{c} x_{c}
\end{equation}
Comparing Figures \ref{fig:bimolecular}(b) and
\ref{fig:bimolecular}(c):
\begin{align}
 \mu_1 &= RT \ln K_{ab} x _{ab} = RT \ln K_{a} x_{a} \label{eq:mu_1}
\end{align}
It follows from Equation (\ref{eq:bimolecular:x_ab}) that:
\begin{equation}
 \label{eq:K_ab}
  K_{ab} = K_a \frac{x_{a}}{x_{ab}} = \frac{K_a}{1 + \tilde{x}_c}
\end{equation}
Using Equations (\ref{eq:mu_2}), (\ref{eq:mu_1}) and (\ref{eq:K_ab})
it follows that:
\begin{align}
 \mu_2 &=  RT \ln K_{ab} x_{ab} + RT \ln K_{c} x_{c}\notag\\
 &= RT \ln K_{ab} x_{ab}  K_{c} x_{c} \notag\\
 &= RT \ln K_{ab} x_{ab}  \frac{K_b}{K_a}\tilde{x}_c \notag\\
 &= RT \ln K_{ba} x_{ab} \label{eq:mu_2_1}\\
\text{where} K_{ba} &= \frac{K_b \tilde{x}_c}{1 + \tilde{x}_c}
\end{align}
Equations (\ref{eq:K_ab}) and (\ref{eq:mu_2_1}) correspond to the
equations for $\alpha_4^+$ and $\alpha_3^-$ in the paper of
\citet[Equations (30) \& (31)]{SmiCra04}.

These simplification approaches for slow-fast reactions can naturally be
applied to more complicated reaction schemes by application in a
stepwise manner, each step of which preserves the underlying
thermodynamic structure of the model. This can be automated, and can
be used to generate different representations of an underlying model,
as for example was done in our recent model of the 
cardiac sarcoplasmic/endoplasmic $\text{Ca}^{2+}$ (SERCA) pump. Such
models consider enzyme mechanisms to be thermodynamic cycles. These
are discussed below.

\section{Example: model reduction of an enzymatic cycle model of the {SERCA} pump}
\label{sec:SERCA}
\begin{figure}[htbp]
 \centering
\SubFig{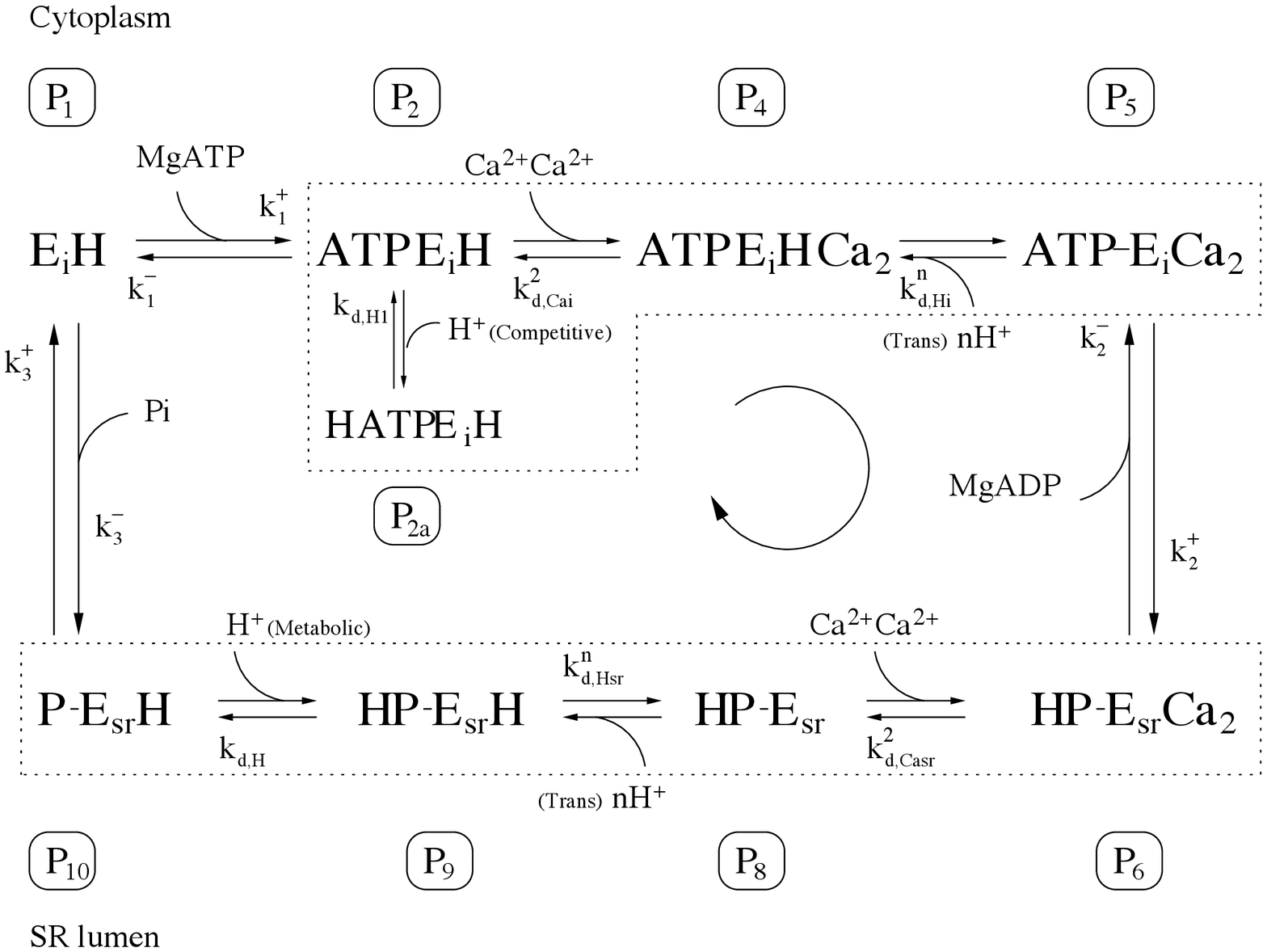}{9-state (8-state cycle) model}{0.6}
\SubFig{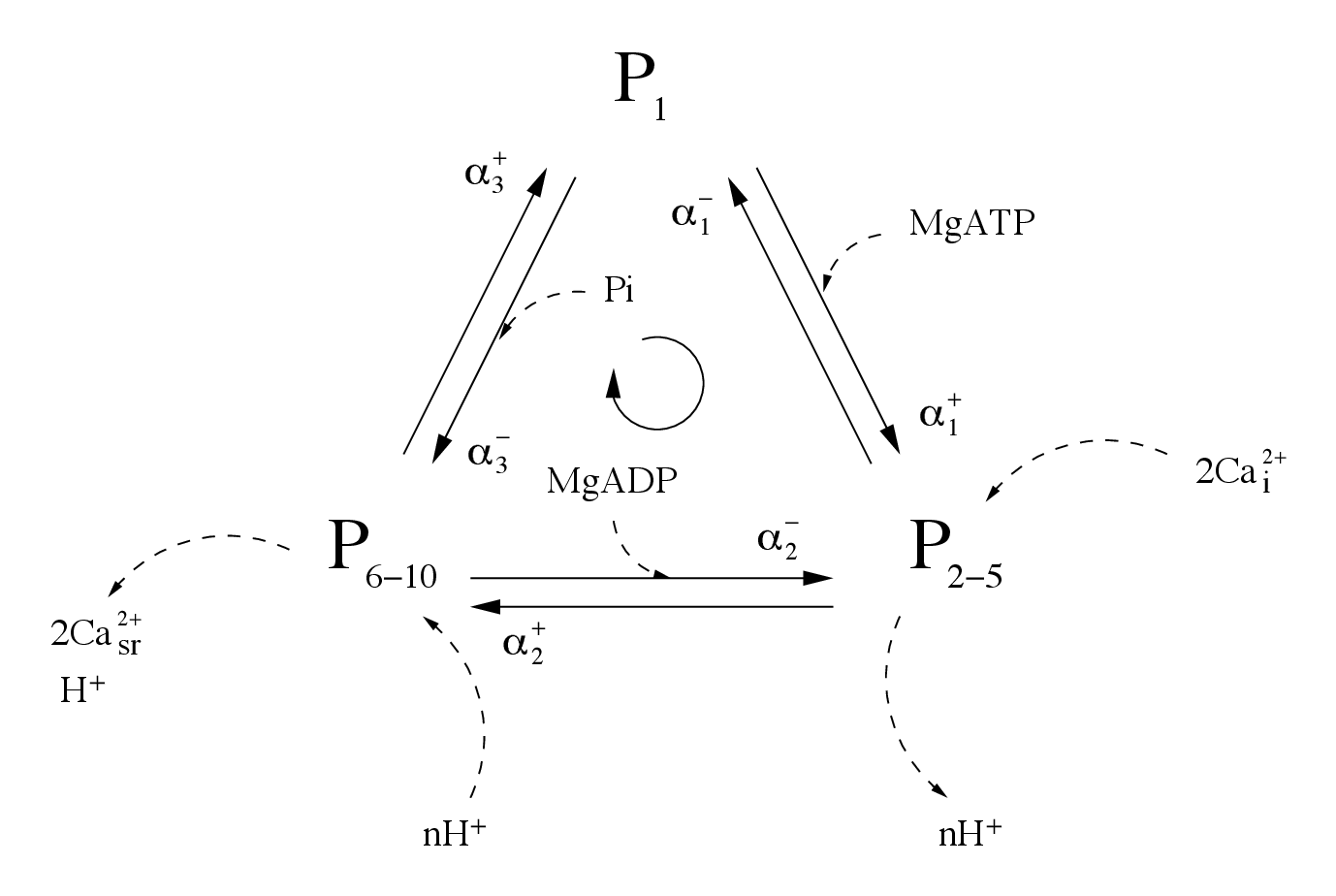}{3-state reduced enzymatic cycle model}{0.6}
 \caption[Schematic of {SERCA} pump model]{Schematic of {SERCA} pump model. (a) for the cardiac SERCA pump (where calcium binding mechanism is assumed to be fully cooperative); and (b) reduced 3-state model, modified from \citet{TraSmiLoiCra09}. The dotted boxes in (a) show partial sub-systems of the model which are simplified to reduce the 9-state to the 3-state model. }
 \label{fig:SERCA}
\end{figure}

\begin{figure}[htbp]
 \centering
 \SubFig{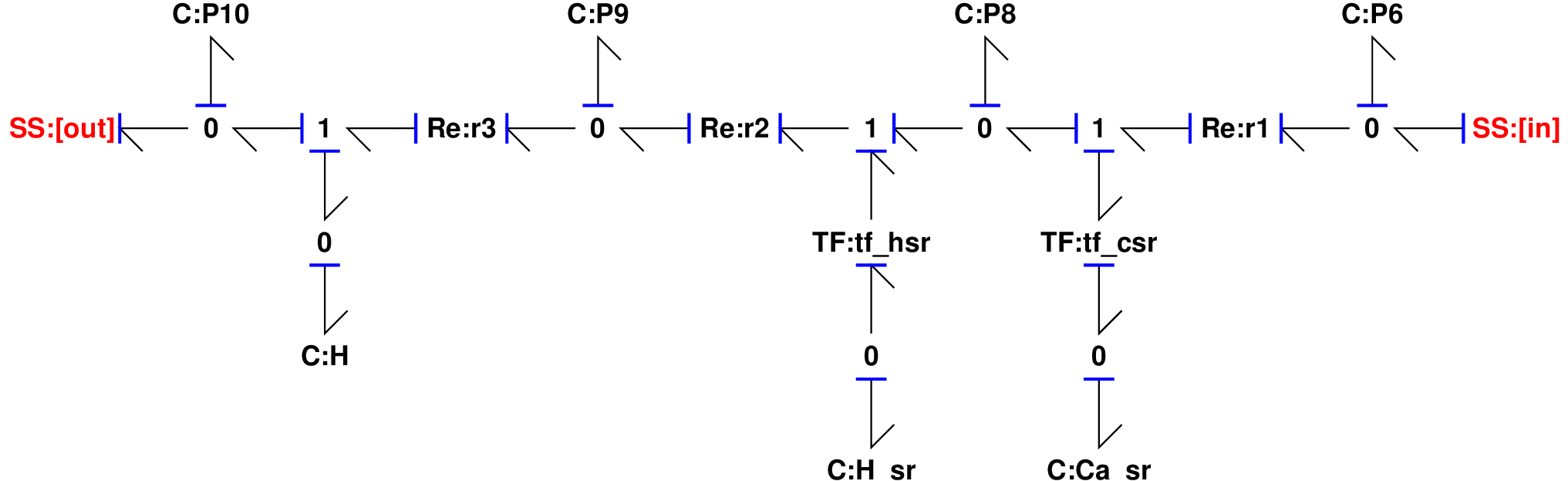}{Bond graph of partial system}{0.8}
 \SubFig{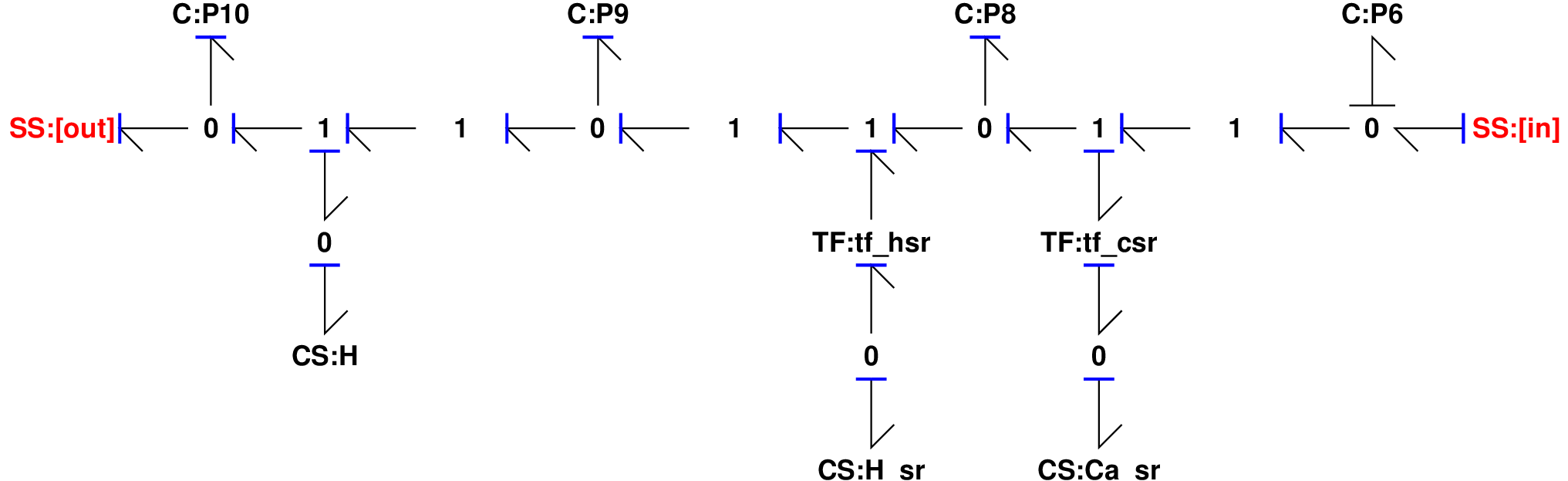}{Bond of simplified partial system}{0.8}
 \SubFig{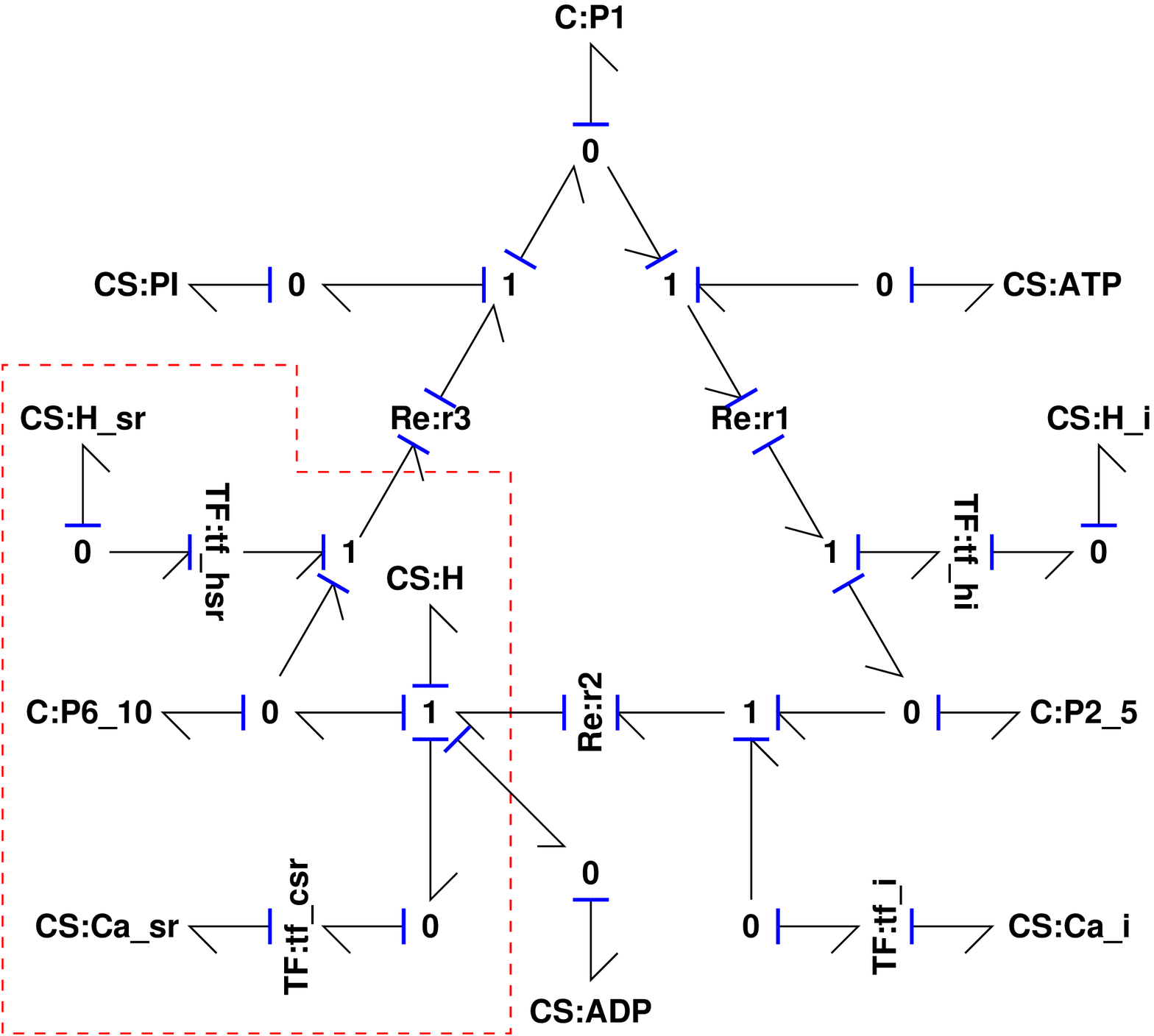}{Bond graph of simplified pump}{0.5}
 \caption[Bond Graph of Partial {SERCA} pump model]{Bond Graph of
   Partial {SERCA} pump model. (a) Shows the bond graph corresponding
   to the subsystem with states $P_6$--$P_{10}$ of Figure
   \ref{subfig:SERCA_Coop_9_State-updated} and the seven \C components
   in integral causality correspond to the seven states. (b) The
   simplified subsystem has one \C component in integral causality and
   thus only one state. (c) This subsystem is shown within the dashed
   box as part of the overall simplified model. \BC{P6\_10}
   represents the composite state of the subsystem and the \CS
   components the corresponding constant concentrations.}
 \label{fig:SERCA_bg}
\end{figure}

\citet{TraSmiLoiCra09} present a thermodynamic enzyme cycle model of the cardiac
sarcoplasmic/endoplasmic $\text{Ca}^{2+}$ ATPase (SERCA) pump. A multi-state model is constructed which incorporates binding of different molecular species to the SERCA protein, including transported calcium ions, co-transported and competitively binding hydrogen ions, ATP and its hydrolysis products ADP, Pi and hydrogen ion, and which represents conformational changes of the protein in the enzymatic cycle, and associated free energy transduction. This generates a thermodynamically constrained enzyme cycle model for SERCA, however the model has a large number of reaction steps and associated parameters, and,
using methods akin to those in Section 4, the model is
reduced from the nine-state, nine-reaction model of Figure
\ref{fig:SERCA}(a) to the three-state, three-reaction model of Figure
\ref{fig:SERCA}(b) by simplifying the reaction mechanism corresponding
to states $P_2$--$P_5$ and to states $P_6$--$P_{10}$ by assuming rapid equilibrium for calcium and hydrogen ion association-dissociation reactions.

To illustrate the bond graph equivalent of this procedure, Figure
\ref{subfig:P610_cbg} gives the reaction mechanism corresponding
to states $P_6$--$P_{10}$. Using the approach of Section
\ref{sec:bimolecular}, this is reduced to the bond graph of Figure
\ref{subfig:P610fast_cbg}. 
The bond graph corresponding to Figure \ref{fig:SERCA}(b) is given in
Figure \ref{subfig:Pump_cbg} where the dotted line delineates the
approximation to reaction mechanism corresponding
to states $P_6$--$P_{10}$.

The three reaction components \BRe{r1}--\BRe{r3} correspond to flows:
\begin{align}
 v_1 &= \alpha_1^+ x_1 - \alpha_1^- x_2\\ 
 v_2 &= \alpha_2^+ x_2 - \alpha_2^- x_3\\ 
 v_3 &= \alpha_3^+ x_3 - \alpha_3^- x_1 
\end{align}
where $x_1$, $x_2$ and $x_3$ are the state occupancy probabilities of states \BC{P_1}, \BC{P_{2-5}}
and \BC{P_{6-10}}. 
Using methods akin to those of Section \ref{sec:bimolecular}, \citet{TraSmiLoiCra09} show that:
\begin{eqnarray}
\alpha^+_1 &=& k^+_1 \mathrm{[MgATP]},  \\
\alpha^+_2 &=& \dfrac{k^+_2 \mathrm{\widetilde{Ca}^2_i}}{\mathrm{\widetilde{Ca}^2_i} (1+ \mathrm{\widetilde{\,H}_i^n}) + \mathrm{\widetilde{\,H}_i^n}(1 + \mathrm{\widetilde{H}_1})},  \\
\alpha^+_3 &=& \dfrac{k^+_3 \mathrm{\widetilde{\,H}_{sr}^n}} {\mathrm{\widetilde{H}}(1 + \mathrm{\widetilde{Ca}^2_{sr}}) + \mathrm{\widetilde{\,H}_{sr}^n}(1 + \mathrm{\widetilde{H}})}
\end{eqnarray}
and the apparent backward rate constants are:
\begin{eqnarray}
\alpha^-_1 &=& \dfrac{k^-_1 \mathrm{\widetilde{\,H}_i^n}}{\mathrm{\widetilde{Ca}^2_i} (1+ \mathrm{\widetilde{\,H}_i^n}) + \mathrm{\widetilde{\,H}_i^n}(1 + \mathrm{\widetilde{H}_1})},  \\
\alpha^-_2 &=& \dfrac{k^-_2 \mathrm{[MgADP]} \mathrm{\widetilde{Ca}^2_{sr}} \mathrm{\widetilde{\,H}_{sr}^n} }{\mathrm{\widetilde{H}}(1 + \mathrm{\widetilde{Ca}^2_{sr}}) + \mathrm{\widetilde{\,H}^n_{sr}}(1 + \mathrm{\widetilde{H}})},     \\
\alpha^-_3 &=& k^-_3 \mathrm{[Pi]}
\end{eqnarray}
where
\begin{eqnarray}
\nonumber
    \mathrm{\widetilde{Ca}_i} = \dfrac{\mathrm{[Ca^{2+}]_i}}{K_{d,Cai}},        & \mathrm{\widetilde{H}_i} = \dfrac{\mathrm{[H^+]}}{K_{d,Hi}},        & \mathrm{\widetilde{H}_1} = \dfrac{\mathrm{[H^+]}}{K_{d,H1}}, \\ \nonumber
    \mathrm{\widetilde{Ca}_{sr}} = \dfrac{\mathrm{[Ca^{2+}]_{sr}}}{K_{d,Casr}},        & \mathrm{\widetilde{H}_{sr}}  = \dfrac{\mathrm{[H^+]}}{K_{d,Hsr}}, & \mathrm{\widetilde{H}} =\dfrac{\mathrm{[H^+]}}{K_{d,H}}
\end{eqnarray}
%
%
%



As discussed further below, the purpose of this model reduction is not to reduce dynamical complexity but rather to reduce the number of unknown parameters to a value consistent with
available experimental data. 
As demonstrated here, and discussed by \citet{TraSmiLoiCra09}, this approach retains the thermodynamic properties of the full enzyme cycle model while reducing the number of unknown parameters. 

The major advantage of this approach is that in constructing models such as this we usually do not know, a priori, the full set of parameters associated with the enzymatic cycle. A subset of the parameters, such as the free energy of hydrolysis of ATP, are known; and these values carry through to the reduced model. However, the majority of parameters (binding and unbinding rates, which are reduced to dissociation constants in the rapid equilibrium approximation) are typically not known and must be estimated by fitting the resulting model to data (namely the steady state cycling rate of the model, as a function of concentrations of the different species, fitted to the data where rate of calcium transport is measured for different concentrations of calcium, pH and metabolites). This parameter estimation process is made significantly more tractable following reduction of the model to the simpler cycle, without compromising the thermodynamic properties and the prior knowledge incorporated in the full multi-state construction. 

\newpage

\begin{thebibliography}{83}
\providecommand{\natexlab}[1]{#1}
\providecommand{\url}[1]{\texttt{#1}}
\expandafter\ifx\csname urlstyle\endcsname\relax
  \providecommand{\doi}[1]{doi: #1}\else
  \providecommand{\doi}{doi: \begingroup \urlstyle{rm}\Url}\fi

\bibitem[Hill(1989)]{Hil89}
Terrell~L Hill.
\newblock \emph{Free energy transduction and biochemical cycle kinetics}.
\newblock Springer-Verlag, New York, 1989.

\bibitem[Beard and Qian(2010)]{BeaQia10}
Daniel~A Beard and Hong Qian.
\newblock \emph{Chemical biophysics: quantitative analysis of cellular
  systems}.
\newblock Cambridge University Press, 2010.

\bibitem[Keener and Sneyd(2009)]{KeeSne09}
James~P Keener and James Sneyd.
\newblock \emph{Mathematical Physiology: {I}: Cellular Physiology}, volume~1.
\newblock Springer, 2nd edition, 2009.

\bibitem[Katchalsky and Curran(1965)]{KatCur65}
A.~Katchalsky and Peter~F. Curran.
\newblock \emph{{Nonequilibrium Thermodynamics in Biophysics}}.
\newblock Harvard University Press, Cambridge, Massachusetts., 1965.

\bibitem[Cellier(1991)]{Cel91}
F.~E. Cellier.
\newblock \emph{Continuous system modelling}.
\newblock Springer-Verlag, 1991.

\bibitem[Gunawardena(2014)]{Gun14}
Jeremy Gunawardena.
\newblock Time-scale separation -- {Michaelis and Menten's} old idea, still
  bearing fruit.
\newblock \emph{FEBS Journal}, 281\penalty0 (2):\penalty0 473--488, 2014.
\newblock ISSN 1742-4658.
\newblock \doi{10.1111/febs.12532}.

\bibitem[Paynter(1992)]{Pay92}
H.M. Paynter.
\newblock An epistemic prehistory of bond graphs.
\newblock In P.C. Breedveld and G.~Dauphin-Tanguy, editors, \emph{Bond Graphs
  for Engineers}, pages 3--17. North-Holland, Amsterdam, 1992.

\bibitem[Gawthrop and Smith(1996)]{GawSmi96}
P.~J. Gawthrop and L.~P.~S. Smith.
\newblock \emph{Metamodelling: Bond Graphs and Dynamic Systems}.
\newblock Prentice Hall, Hemel Hempstead, Herts, England., 1996.
\newblock ISBN 0-13-489824-9.

\bibitem[Borutzky(2011)]{Bor11}
Wolfgang Borutzky.
\newblock \emph{Bond Graph Modelling of Engineering Systems: Theory,
  Applications and Software Support}.
\newblock Springer, 2011.
\newblock ISBN 9781441993670.

\bibitem[Karnopp et~al.(2012)Karnopp, Margolis, and Rosenberg]{KarMarRos12}
Dean~C Karnopp, Donald~L Margolis, and Ronald~C Rosenberg.
\newblock \emph{System Dynamics: Modeling, Simulation, and Control of
  Mechatronic Systems}.
\newblock John Wiley \& Sons, 5th edition, 2012.
\newblock ISBN 978-0470889084.

\bibitem[Gawthrop and Bevan(2007)]{GawBev07}
Peter~J Gawthrop and Geraint~P Bevan.
\newblock Bond-graph modeling: A tutorial introduction for control engineers.
\newblock \emph{IEEE Control Systems Magazine}, 27\penalty0 (2):\penalty0
  24--45, April 2007.
\newblock \doi{10.1109/MCS.2007.338279}.

\bibitem[Palsson(2006)]{Pal06}
Bernhard Palsson.
\newblock \emph{Systems biology: properties of reconstructed networks}.
\newblock Cambridge University Press, 2006.
\newblock ISBN 0521859034.

\bibitem[Palsson(2011)]{Pal11}
Bernhard Palsson.
\newblock \emph{Systems Biology: Simulation of Dynamic Network States}.
\newblock Cambridge University Press, 2011.

\bibitem[Alon(2007)]{Alo07}
Uri Alon.
\newblock \emph{Introduction to Systems Biology: Design Principles of
  Biological Networks}.
\newblock CRC press, 2007.

\bibitem[Klipp et~al.(2011)Klipp, Liebermeister, Wierling, Kowald, Lehrach, and
  Herwig]{KliLieWie11}
Edda Klipp, Wolfram Liebermeister, Christoph Wierling, Axel Kowald, Hans
  Lehrach, and Ralf Herwig.
\newblock \emph{Systems biology}.
\newblock Wiley-Blackwell, 2011.

\bibitem[Oster et~al.(1971)Oster, Perelson, and Katchalsky]{OstPerKat71}
George Oster, Alan Perelson, and Aharon Katchalsky.
\newblock Network thermodynamics.
\newblock \emph{Nature}, 234:\penalty0 393--399, December 1971.
\newblock \doi{10.1038/234393a0}.

\bibitem[Oster et~al.(1973)Oster, Perelson, and Katchalsky]{OstPerKat73}
George~F. Oster, Alan~S. Perelson, and Aharon Katchalsky.
\newblock Network thermodynamics: dynamic modelling of biophysical systems.
\newblock \emph{Quarterly Reviews of Biophysics}, 6\penalty0 (01):\penalty0
  1--134, 1973.
\newblock \doi{10.1017/S0033583500000081}.

\bibitem[Oster and Perelson(1974)]{OstPer74}
G.~Oster and A.~Perelson.
\newblock Chemical reaction networks.
\newblock \emph{Circuits and Systems, IEEE Transactions on}, 21\penalty0
  (6):\penalty0 709 -- 721, November 1974.
\newblock ISSN 0098-4094.
\newblock \doi{10.1109/TCS.1974.1083946}.

\bibitem[Oster and Auslander(1971{\natexlab{a}})]{OstAus71a}
George~F. Oster and David~M. Auslander.
\newblock Topological representations of thermodynamic systems--{I}. basic
  concepts.
\newblock \emph{Journal of the Franklin Institute}, 292\penalty0 (1):\penalty0
  1 -- 17, 1971{\natexlab{a}}.
\newblock ISSN 0016-0032.
\newblock \doi{10.1016/0016-0032(71)90037-8}.

\bibitem[Oster and Auslander(1971{\natexlab{b}})]{OstAus71b}
George~F. Oster and David~M. Auslander.
\newblock Topological representations of thermodynamic systems--{II}. some
  elemental subunits for irreversible thermodynamics.
\newblock \emph{Journal of the Franklin Institute}, 292\penalty0 (2):\penalty0
  77 -- 92, 1971{\natexlab{b}}.
\newblock ISSN 0016-0032.
\newblock \doi{10.1016/0016-0032(71)90196-7}.

\bibitem[Kohl et~al.(2010)Kohl, Crampin, Quinn, and Noble]{Kohl:2010iw}
P~Kohl, Edmund~J Crampin, T~A Quinn, and D~Noble.
\newblock {Systems Biology: An Approach}.
\newblock \emph{Clinical Pharmacology {\&} Therapeutics}, 88\penalty0
  (1):\penalty0 25--33, June 2010.

\bibitem[Aldridge et~al.(2006)Aldridge, Burke, Lauffenburger, and
  Sorger]{AldBurLauSor06}
Bree~B. Aldridge, John~M. Burke, Douglas~A. Lauffenburger, and Peter~K. Sorger.
\newblock Physicochemical modelling of cell signalling pathways.
\newblock \emph{Nat Cell Biol}, 8:\penalty0 1195--1203, November 2006.
\newblock ISSN 1465-7392.
\newblock \doi{10.1038/ncb1497}.

\bibitem[Smith et~al.(2007)Smith, Crampin, Niederer, Bassingthwaighte, and
  Beard]{Smith:2007co}
Nicolas~P Smith, Edmund~J Crampin, Steven~A Niederer, James~B Bassingthwaighte,
  and Daniel~A Beard.
\newblock {Computational biology of cardiac myocytes: proposed standards for
  the physiome.}
\newblock \emph{The Journal of experimental biology}, 210\penalty0 (Pt
  9):\penalty0 1576--1583, May 2007.

\bibitem[Hunter et~al.(2008)Hunter, Crampin, and Nielsen]{Hunter:2008cb}
Peter~J Hunter, Edmund~J Crampin, and Poul M~F Nielsen.
\newblock {Bioinformatics, multiscale modeling and the IUPS Physiome Project.}
\newblock \emph{Briefings in Bioinformatics}, 9\penalty0 (4):\penalty0
  333--343, July 2008.

\bibitem[Hunter et~al.(2012)Hunter, Bradley, Britten, Brooks, Carotenuto,
  Christie, Frangi, Garny, Ladd, Caton~Little, Nielsen, Miller, Planes,
  Steghoffer, Young, and Yu]{Hunter:2012dd}
Peter Hunter, Chris Bradley, Randall Britten, David Brooks, Luigi Carotenuto,
  Richard Christie, Alejandro Frangi, Alan Garny, David Ladd, David~Nickerson
  Caton~Little, Poul Nielsen, Andrew Miller, Xavier Planes, Martin Steghoffer,
  Alistair Young, and Tommy Yu.
\newblock {The VPH-Physiome Project: standards, tools and databases for
  multi-scale physiological modelling}.
\newblock In Ambrosi D, Quarteroni A, and Rozza G, editors, \emph{Modeling of
  Physiological Flows.}, pages 1--23. Springer-Verlag Italia, November 2012.

\bibitem[Smith and Crampin(2004)]{SmiCra04}
N.P. Smith and E.J. Crampin.
\newblock Development of models of active ion transport for whole-cell
  modelling: cardiac sodium-potassium pump as a case study.
\newblock \emph{Progress in Biophysics and Molecular Biology}, 85\penalty0
  (2-3):\penalty0 387 -- 405, 2004.
\newblock \doi{10.1016/j.pbiomolbio.2004.01.010}.

\bibitem[Tran et~al.(2009{\natexlab{a}})Tran, Smith, Loiselle, and
  Crampin]{Tran:2009ui}
Kenneth Tran, Nicolas~P Smith, Denis~S Loiselle, and Edmund~J Crampin.
\newblock {A Thermodynamic Model of the Cardiac Sarcoplasmic/Endoplasmic Ca2+
  (SERCA) Pump}.
\newblock \emph{Biophysical Journal}, 96\penalty0 (5):\penalty0 2029--2042,
  2009{\natexlab{a}}.

\bibitem[Beard(2005)]{Beard:2005uk}
Daniel~A Beard.
\newblock {A biophysical model of the mitochondrial respiratory system and
  oxidative phosphorylation}.
\newblock \emph{PLoS Computational Biology}, 1\penalty0 (4):\penalty0 e36,
  2005.

\bibitem[Beard et~al.(2004)Beard, Babson, Curtis, and Qian]{Beard:2004iu}
Daniel~A Beard, Eric Babson, Edward Curtis, and Hong Qian.
\newblock {Thermodynamic constraints for biochemical networks}.
\newblock \emph{Journal of Theoretical Biology}, 228\penalty0 (3):\penalty0
  327--333, June 2004.

\bibitem[Beard et~al.(2002)Beard, Liang, and Qian]{BeaLiaQia02}
Daniel~A. Beard, Shoudan Liang, and Hong Qian.
\newblock Energy balance for analysis of complex metabolic networks.
\newblock \emph{Biophysical Journal}, 83\penalty0 (1):\penalty0 79 -- 86, 2002.
\newblock ISSN 0006-3495.
\newblock \doi{10.1016/S0006-3495(02)75150-3}.

\bibitem[Feist et~al.(2007)Feist, Henry, Reed, Krummenacker, Joyce, Karp,
  Broadbelt, Hatzimanikatis, and Palsson]{Feist:2007dq}
Adam~M Feist, Christopher~S Henry, Jennifer~L Reed, Markus Krummenacker,
  Andrew~R Joyce, Peter~D Karp, Linda~J Broadbelt, Vassily Hatzimanikatis, and
  Bernhard~O Palsson.
\newblock {A genome-scale metabolic reconstruction for Escherichia coli K-12
  MG1655 that accounts for 1260 ORFs and thermodynamic information}.
\newblock \emph{Molecular Systems Biology}, 3, June 2007.

\bibitem[Soh and Hatzimanikatis(2010)]{Soh:2010je}
Keng~Cher Soh and Vassily Hatzimanikatis.
\newblock {Network thermodynamics in the post-genomic era.}
\newblock \emph{Current Opinion in Microbiology}, 13\penalty0 (3):\penalty0
  350--357, June 2010.

\bibitem[Ballance et~al.(2005)Ballance, Bevan, Gawthrop, and
  Diston]{BalBevGawDis05}
Donald~J. Ballance, Geraint~P. Bevan, Peter~J. Gawthrop, and Dominic~J. Diston.
\newblock Model transformation tools ({MTT}): The open source bond graph
  project.
\newblock In \emph{Proceedings of the 2005 International Conference On Bond
  Graph Modeling and Simulation (ICBGM'05)}, Simulation Series, pages 123--128,
  New Orleans, U.S.A., January 2005. Society for Computer Simulation.

\bibitem[Cellier and Nebot(2005)]{CelNeb05}
F.E. Cellier and A.~Nebot.
\newblock The modelica bond graph library.
\newblock In \emph{Proceedings 4th International Modelica Conference},
  volume~1, pages 57--65, Hamburg, Germany, 2005.

\bibitem[Borutzky(2006)]{Bor06}
W.~Borutzky.
\newblock {BGML} -- a novel {XML} format for the exchange and the reuse of bond
  graph models of engineering systems.
\newblock \emph{Simulation Modelling Practice and Theory}, 14\penalty0
  (7):\penalty0 787 -- 808, 2006.
\newblock ISSN 1569-190X.
\newblock \doi{10.1016/j.simpat.2006.01.002}.

\bibitem[Cellier and Greifeneder(2008)]{CelGre08}
F.E. Cellier and J.~Greifeneder.
\newblock Thermobondlib - a new modelica library for modeling convective flows.
\newblock In \emph{Proceedings of the 6th International Modelica Conference},
  pages 163 -- 172, Bielefeld, Deutschland, March 2008.

\bibitem[Cellier and Greifeneder(2009)]{CelGre09}
F.E. Cellier and J.~Greifeneder.
\newblock Modeling chemical reactions in modelica by use of chemo-bonds.
\newblock In \emph{Proceedings 7th Modelica Conference}, Como, Italy, September
  2009.

\bibitem[de~la Calle et~al.(2013)de~la Calle, Cellier, Yebra, and
  Dormido]{CalCelYebDor13}
Alberto de~la Calle, Francois~E. Cellier, Luis~J. Yebra, and Sebastian Dormido.
\newblock Improvements in bondlib the modelica bond graph library.
\newblock In \emph{Proceeding of the 8th EUROSIM Congress, Cardiff}, Cardiff,
  Wales, September 2013.

\bibitem[Karnopp(1990)]{Kar90}
Dean Karnopp.
\newblock Bond graph models for electrochemical energy storage : electrical,
  chemical and thermal effects.
\newblock \emph{Journal of the Franklin Institute}, 327\penalty0 (6):\penalty0
  983 -- 992, 1990.
\newblock ISSN 0016-0032.
\newblock \doi{10.1016/0016-0032(90)90073-R}.

\bibitem[Thoma and Atlan(1977)]{ThoAtl77}
Jean~U. Thoma and Henri Atlan.
\newblock Network thermodynamics with entropy stripping.
\newblock \emph{Journal of the Franklin Institute}, 303\penalty0 (4):\penalty0
  319 -- 328, 1977.
\newblock ISSN 0016-0032.
\newblock \doi{10.1016/0016-0032(77)90114-4}.

\bibitem[Greifeneder and Cellier(2012)]{GreCel12}
J.~Greifeneder and F.E. Cellier.
\newblock Modeling chemical reactions using bond graphs.
\newblock In \emph{Proceedings ICBGM12, 10th SCS Intl. Conf. on Bond Graph
  Modeling and Simulation}, pages 110--121, Genoa, Italy, 2012.

\bibitem[Thoma and Atlan(1985)]{ThoAtl85}
Jean Thoma and Henri Atlan.
\newblock Osmosis and hydraulics by network thermodynamics and bond graphs.
\newblock \emph{Journal of the Franklin Institute}, 319\penalty0
  (1-2):\penalty0 217 -- 226, 1985.
\newblock ISSN 0016-0032.
\newblock \doi{10.1016/0016-0032(85)90075-4}.

\bibitem[LeF\`{e}vre et~al.(1999)LeF\`{e}vre, LeF\`{e}vre, and
  Couteiro]{LefLefCou99}
Jacques LeF\`{e}vre, Laurent LeF\`{e}vre, and Bernadette Couteiro.
\newblock A bond graph model of chemo-mechanical transduction in the mammalian
  left ventricle.
\newblock \emph{Simulation Practice and Theory}, 7\penalty0 (5-6):\penalty0
  531--552, 1999.
\newblock ISSN 0928-4869.
\newblock \doi{10.1016/S0928-4869(99)00023-3}.

\bibitem[Fuchs(1996)]{Fuc96}
Hans~U. Fuchs.
\newblock \emph{The Dynamics of Heat}.
\newblock Springer, New York, 1996.

\bibitem[Job and Herrmann(2006)]{JobHer06}
G~Job and F~Herrmann.
\newblock Chemical potential -- a quantity in search of recognition.
\newblock \emph{European Journal of Physics}, 27\penalty0 (2):\penalty0
  353--371, 2006.
\newblock \doi{10.1088/0143-0807/27/2/018}.

\bibitem[Cellier(1992)]{Cel92}
F.~E. Cellier.
\newblock Hierarchical non-linear bond graphs: a unified methodology for
  modeling complex physical systems.
\newblock \emph{SIMULATION}, 58\penalty0 (4):\penalty0 230--248, 1992.
\newblock \doi{10.1177/003754979205800404}.

\bibitem[Gawthrop and Smith(1992)]{GawSmi92a}
P.~J. Gawthrop and L.~Smith.
\newblock Causal augmentation of bond graphs with algebraic loops.
\newblock \emph{Journal of the Franklin Institute}, 329\penalty0 (2):\penalty0
  291--303, 1992.
\newblock \doi{10.1016/0016-0032(92)90035-F}.

\bibitem[Sueur and Dauphin-Tanguy(1991)]{SueDau91a}
C.~Sueur and G.~Dauphin-Tanguy.
\newblock Bond graph approach to multi-time scale systems analysis.
\newblock \emph{Journal of the Franklin Institute}, 328\penalty0
  (5•¡¹6):\penalty0 1005 -- 1026, 1991.
\newblock ISSN 0016-0032.
\newblock \doi{10.1016/0016-0032(91)90066-C}.

\bibitem[Qian and Beard(2005)]{QiaBea05}
Hong Qian and Daniel~A. Beard.
\newblock Thermodynamics of stoichiometric biochemical networks in living
  systems far from equilibrium.
\newblock \emph{Biophysical Chemistry}, 114\penalty0 (2-3):\penalty0 213 --
  220, 2005.
\newblock ISSN 0301-4622.
\newblock \doi{10.1016/j.bpc.2004.12.001}.

\bibitem[Qian et~al.(2003)Qian, Beard, and Liang]{QiaBeaLia03}
Hong Qian, Daniel~A. Beard, and Shou-dan Liang.
\newblock Stoichiometric network theory for nonequilibrium biochemical systems.
\newblock \emph{European Journal of Biochemistry}, 270\penalty0 (3):\penalty0
  415--421, 2003.
\newblock ISSN 1432-1033.
\newblock \doi{10.1046/j.1432-1033.2003.03357.x}.

\bibitem[Van~Rysselberghe(1958)]{Rys58}
Pierre Van~Rysselberghe.
\newblock Reaction rates and affinities.
\newblock \emph{The Journal of Chemical Physics}, 29\penalty0 (3):\penalty0
  640--642, 1958.
\newblock \doi{10.1063/1.1744552}.

\bibitem[Boudart(1983)]{Bou83}
M.~Boudart.
\newblock Thermodynamic and kinetic coupling of chain and catalytic reactions.
\newblock \emph{The Journal of Physical Chemistry}, 87\penalty0 (15):\penalty0
  2786--2789, 1983.
\newblock \doi{10.1021/j100238a018}.

\bibitem[Laidler(1985)]{Lai85}
Keith~J. Laidler.
\newblock {Ren\'{e} Marcelin} (1885-1914), a short-lived genius of chemical
  kinetics.
\newblock \emph{Journal of Chemical Education}, 62\penalty0 (11):\penalty0
  1012, 1985.
\newblock \doi{10.1021/ed062p1012}.

\bibitem[Jamshidi and Palsson(2011)]{JamPal11}
Neema Jamshidi and Bernhard Palsson.
\newblock Metabolic network dynamics: Properties and principles.
\newblock In Jennifer~Southgate Werner~Dubitzky and Hendrik Fuss, editors,
  \emph{Understanding the Dynamics of Biological Systems}, pages 19--37.
  Springer, Berlin, 2011.
\newblock \doi{10.1007/978-1-4419-7964-3\_2}.

\bibitem[Schilling et~al.(2000)Schilling, Letscher, and Palsson]{SchLetPal00}
Christophe~H. Schilling, David Letscher, and Bernhard Palsson.
\newblock Theory for the systemic definition of metabolic pathways and their
  use in interpreting metabolic function from a pathway-oriented perspective.
\newblock \emph{Journal of Theoretical Biology}, 203\penalty0 (3):\penalty0 229
  -- 248, 2000.
\newblock ISSN 0022-5193.
\newblock \doi{10.1006/jtbi.2000.1073}.

\bibitem[Schuster et~al.(2002)Schuster, Hilgetag, Woods, and
  Fell]{SchHilWooFel02}
S.~Schuster, C.~Hilgetag, J.H. Woods, and D.A. Fell.
\newblock Reaction routes in biochemical reaction systems: Algebraic
  properties, validated calculation procedure and example from nucleotide
  metabolism.
\newblock \emph{Journal of Mathematical Biology}, 45\penalty0 (2):\penalty0
  153--181, 2002.
\newblock ISSN 0303-6812.
\newblock \doi{10.1007/s002850200143}.

\bibitem[Famili and Palsson(2003{\natexlab{a}})]{FamPal03}
Iman Famili and Bernhard~O. Palsson.
\newblock Systemic metabolic reactions are obtained by singular value
  decomposition of genome-scale stoichiometric matrices.
\newblock \emph{Journal of Theoretical Biology}, 224\penalty0 (1):\penalty0 87
  -- 96, 2003{\natexlab{a}}.
\newblock ISSN 0022-5193.
\newblock \doi{10.1016/S0022-5193(03)00146-2}.

\bibitem[Famili and Palsson(2003{\natexlab{b}})]{FamPal03a}
Iman Famili and Bernhard~O. Palsson.
\newblock The convex basis of the left null space of the stoichiometric matrix
  leads to the definition of metabolically meaningful pools.
\newblock \emph{Biophysical Journal}, 85\penalty0 (1):\penalty0 16 -- 26,
  2003{\natexlab{b}}.
\newblock ISSN 0006-3495.
\newblock \doi{10.1016/S0006-3495(03)74450-6}.

\bibitem[Gawthrop(2000{\natexlab{a}})]{Gaw00d}
Peter~J Gawthrop.
\newblock Physical interpretation of inverse dynamics using bicausal bond
  graphs.
\newblock \emph{Journal of the Franklin Institute}, 337\penalty0 (6):\penalty0
  743--769, 2000{\natexlab{a}}.
\newblock \doi{10.1016/S0016-0032(00)00051-X}.

\bibitem[Ngwompo et~al.(2001)Ngwompo, Scavarda, and Thomasset]{NgwScaTho01b}
R.~Ngwompo, S.~Scavarda, and D.~Thomasset.
\newblock Physical model-based inversion in control systems design using bond
  graph representation part 2: applications.
\newblock \emph{Proceedings of the I MECH E Part I Journal of Systems and
  Control Engineering}, 215\penalty0 (2):\penalty0 105--112, April 2001.

\bibitem[Marquis-Favre and Jardin(2011)]{MarJar11}
Wilfrid Marquis-Favre and Audrey Jardin.
\newblock Bond graphs and inverse modeling for mechatronic system design.
\newblock In Wolfgang Borutzky, editor, \emph{Bond Graph Modelling of
  Engineering Systems}, pages 195--226. Springer New York, 2011.
\newblock ISBN 978-1-4419-9368-7.
\newblock \doi{10.1007/978-1-4419-9368-7-6}.

\bibitem[Sueur and Dauphin-Tanguy(1989)]{SueDau89}
C.~Sueur and G.~Dauphin-Tanguy.
\newblock Structural controllability/observability of linear systems
  represented by bond graphs.
\newblock \emph{Journal of the Franklin Institute}, 326:\penalty0 869--883,
  1989.

\bibitem[Reder(1988)]{Red88}
Christine Reder.
\newblock Metabolic control theory: A structural approach.
\newblock \emph{Journal of Theoretical Biology}, 135\penalty0 (2):\penalty0 175
  -- 201, 1988.
\newblock ISSN 0022-5193.
\newblock \doi{10.1016/S0022-5193(88)80073-0}.

\bibitem[Ingalls and Sauro(2003)]{IngSau03}
Brian~P. Ingalls and Herbert~M. Sauro.
\newblock Sensitivity analysis of stoichiometric networks: an extension of
  metabolic control analysis to non-steady state trajectories.
\newblock \emph{Journal of Theoretical Biology}, 222\penalty0 (1):\penalty0 23
  -- 36, 2003.
\newblock ISSN 0022-5193.
\newblock \doi{10.1016/S0022-5193(03)00011-0}.

\bibitem[Ingalls(2004)]{Ing04}
Brian~P. Ingalls.
\newblock A frequency domain approach to sensitivity analysis of biochemical
  networks.
\newblock \emph{The Journal of Physical Chemistry B}, 108\penalty0
  (3):\penalty0 1143--1152, 2004.
\newblock \doi{10.1021/jp036567u}.

\bibitem[Sauro(2009)]{Sau09}
H.M. Sauro.
\newblock Network dynamics.
\newblock In Reneé Ireton, Kristina Montgomery, Roger Bumgarner, Ram
  Samudrala, and Jason McDermott, editors, \emph{Computational Systems
  Biology}, volume 541 of \emph{Methods in Molecular Biology}, pages 269--309.
  Humana Press, 2009.
\newblock ISBN 978-1-58829-905-5.
\newblock \doi{10.1007/978-1-59745-243-4\_13}.

\bibitem[Ingalls(2013)]{Ing13}
Brian~P. Ingalls.
\newblock \emph{Mathematical Modelling in Systems Biology}.
\newblock MIT Press, 2013.

\bibitem[Kraeutler et~al.(2010)Kraeutler, Soltis, and Saucerman]{KraSolSau10}
Matthew Kraeutler, Anthony Soltis, and Jeffrey Saucerman.
\newblock Modeling cardiac beta-adrenergic signaling with normalized-hill
  differential equations: comparison with a biochemical model.
\newblock \emph{BMC Systems Biology}, 4\penalty0 (1):\penalty0 157, 2010.
\newblock ISSN 1752-0509.
\newblock \doi{10.1186/1752-0509-4-157}.

\bibitem[Ryall et~al.(2012)Ryall, Holland, Delaney, Kraeutler, Parker, and
  Saucerman]{RyaHolDel12}
Karen~A. Ryall, David~O. Holland, Kyle~A. Delaney, Matthew~J. Kraeutler,
  Audrey~J. Parker, and Jeffrey~J. Saucerman.
\newblock Network reconstruction and systems analysis of cardiac myocyte
  hypertrophy signaling.
\newblock \emph{Journal of Biological Chemistry}, 287\penalty0 (50):\penalty0
  42259--42268, 2012.
\newblock \doi{10.1074/jbc.M112.382937}.

\bibitem[Tran et~al.(2009{\natexlab{b}})Tran, Smith, Loiselle, and
  Crampin]{TraSmiLoiCra09}
Kenneth Tran, Nicolas~P. Smith, Denis~S. Loiselle, and Edmund~J. Crampin.
\newblock A thermodynamic model of the cardiac sarcoplasmic/endoplasmic {Ca2+
  (SERCA)} pump.
\newblock \emph{Biophysical Journal}, 96\penalty0 (5):\penalty0 2029 -- 2042,
  2009{\natexlab{b}}.
\newblock ISSN 0006-3495.
\newblock \doi{10.1016/j.bpj.2008.11.045}.

\bibitem[Fink et~al.(2011)Fink, Niederer, Cherry, Fenton, Koivumaki, Seemann,
  Thul, Zhang, Sachse, Beard, Crampin, and Smith]{FinNieCra11}
Martin Fink, Steven~A. Niederer, Elizabeth~M. Cherry, Flavio~H. Fenton,
  Jussi~T. Koivumaki, Gunnar Seemann, Rudiger Thul, Henggui Zhang, Frank~B.
  Sachse, Dan Beard, Edmund~J. Crampin, and Nicolas~P. Smith.
\newblock Cardiac cell modelling: Observations from the heart of the cardiac
  physiome project.
\newblock \emph{Progress in Biophysics and Molecular Biology}, 104\penalty0
  (1-3):\penalty0 2 -- 21, 2011.
\newblock ISSN 0079-6107.
\newblock \doi{10.1016/j.pbiomolbio.2010.03.002}.

\bibitem[Lloyd et~al.(2004)Lloyd, Halstead, and Nielsen]{Lloyd:2004cv}
Catherine~M Lloyd, Matt~DB Halstead, and Poul~F Nielsen.
\newblock {CellML: its future, present and past}.
\newblock \emph{Progress in Biophysics and Molecular Biology}, 85\penalty0
  (2):\penalty0 433--450, 2004.

\bibitem[Hucka et~al.(2003)Hucka, Finney, Sauro, Bolouri, Doyle, Kitano, Arkin,
  Bornstein, Bray, Cornish-Bowden, Cuellar, Dronov, Gilles, Ginkel, Gor,
  Goryanin, Hedley, Hodgman, Hofmeyr, Hunter, Juty, Kasberger, Kremling,
  Kummer, Le~Nov{\`e}re, Loew, Lucio, Mendes, Minch, Mjolsness, Nakayama,
  Nelson, Nielsen, Sakurada, Schaff, Shapiro, Shimizu, Spence, Stelling,
  Takahashi, Tomita, Wagner, and Wang]{Hucka:2003fs}
M~Hucka, A~Finney, H~M Sauro, H~Bolouri, J~C Doyle, H~Kitano, A~P Arkin, B~J
  Bornstein, D~Bray, A~Cornish-Bowden, A~A Cuellar, S~Dronov, E~D Gilles,
  M~Ginkel, V~Gor, I~I Goryanin, W~J Hedley, T~C Hodgman, J~H Hofmeyr, P~J
  Hunter, N~S Juty, J~L Kasberger, A~Kremling, U~Kummer, N~Le~Nov{\`e}re, L~M
  Loew, D~Lucio, P~Mendes, E~Minch, E~D Mjolsness, Y~Nakayama, M~R Nelson, P~F
  Nielsen, T~Sakurada, J~C Schaff, B~E Shapiro, T~S Shimizu, H~D Spence,
  J~Stelling, K~Takahashi, M~Tomita, J~Wagner, and J~Wang.
\newblock {The systems biology markup language (SBML): a medium for
  representation and exchange of biochemical network models}.
\newblock \emph{Bioinformatics}, 19\penalty0 (4):\penalty0 524--531, March
  2003.

\bibitem[Cooling et~al.(2008)Cooling, Hunter, and Crampin]{CooHunCra08}
M.T. Cooling, P.~Hunter, and E.J. Crampin.
\newblock Modelling biological modularity with {CellML}.
\newblock \emph{Systems Biology, IET}, 2\penalty0 (2):\penalty0 73 --79, March
  2008.
\newblock ISSN 1751-8849.
\newblock \doi{10.1049/iet-syb:20070020}.

\bibitem[Frank(1978)]{Fra78}
Paul~M. Frank.
\newblock \emph{Introduction to System Sensitivity Theory}.
\newblock Academic Press, New York, 1978.

\bibitem[Rosenwasser and Yusupov(2000)]{RosYus00}
Efim Rosenwasser and Rafael Yusupov.
\newblock \emph{Sensitivity of Automatic Control Systems}.
\newblock {CRC} press, Boca Raton, 2000.

\bibitem[Heinrich and Schuster(1996)]{HeiSch96}
Reinhart Heinrich and Stefan Schuster.
\newblock \emph{The regulation of cellular systems}.
\newblock Chapman \& Hall New York, 1996.

\bibitem[Karnopp(1977)]{Kar77}
Dean Karnopp.
\newblock Power and energy in linearized physical systems.
\newblock \emph{Journal of the Franklin Institute}, 303\penalty0 (1):\penalty0
  85 -- 98, 1977.
\newblock ISSN 0016-0032.
\newblock \doi{10.1016/0016-0032(77)90078-3}.

\bibitem[Gawthrop(2000{\natexlab{b}})]{Gaw00c}
Peter~J Gawthrop.
\newblock Sensitivity bond graphs.
\newblock \emph{Journal of the Franklin Institute}, 337\penalty0 (7):\penalty0
  907--922, November 2000{\natexlab{b}}.
\newblock \doi{10.1016/S0016-0032(00)00052-1}.

\bibitem[Gawthrop and Ronco(2000)]{GawRon00b}
Peter~J. Gawthrop and Eric Ronco.
\newblock Estimation and control of mechatronic systems using sensitivity bond
  graphs.
\newblock \emph{Control Engineering Practice}, 8\penalty0 (11):\penalty0
  1237--1248, November 2000.
\newblock \doi{10.1016/S0967-0661(00)00062-9}.

\bibitem[Sueur and Dauphin-Tanguy(1997)]{SueDau97}
C.~Sueur and G.~Dauphin-Tanguy.
\newblock Controllability indices for structured systems.
\newblock \emph{Linear Algebra and its Applications}, 250:\penalty0 275--287,
  1997.

\bibitem[Ngwompo et~al.(1996)Ngwompo, Scavarda, and Thomasset]{NgwScaTho96}
R.~Fotsu Ngwompo, S.~Scavarda, and D.~Thomasset.
\newblock Inversion of linear time-invariant siso systems modelled by bond
  graph.
\newblock \emph{Journal of the Franklin Institute}, 333:\penalty0 157--174,
  March 1996.

\bibitem[Ngwompo and Gawthrop(1999)]{NgwGaw99}
Roger~F Ngwompo and Peter~J Gawthrop.
\newblock Bond graph based simulation of nonlinear inverse systems using
  physical performance specifications.
\newblock \emph{Journal of the Franklin Institute}, 336\penalty0 (8):\penalty0
  1225--1247, November 1999.
\newblock \doi{10.1016/S0016-0032(99)00032-0}.

\end{thebibliography}

\end{document}